\documentclass[a4paper,fleqn]{cas-dc}

\usepackage{cite}
\usepackage{enumerate}
\usepackage[linesnumbered,ruled,vlined]{algorithm2e}
\usepackage[utf8]{inputenc}
\usepackage{graphicx}
\usepackage{subcaption}
\graphicspath{ {./fig/} }
\usepackage{textcomp}
\usepackage{url}
\usepackage[table]{xcolor}
\usepackage{booktabs}
\usepackage{multirow}
\usepackage{xcolor}
\definecolor{linkblue}{RGB}{33,150,209}

\usepackage{amsmath}
\usepackage{float}
\usepackage[noend]{algpseudocode}
\usepackage[]{hyperref}
\hypersetup{
        colorlinks = true,
        citecolor=linkblue,
        linkcolor=linkblue,
        urlcolor=linkblue
}

\usepackage{tabularray}
\newcommand{\JK}[1]{\textcolor{black}{#1}}
\newcommand{\R}[1]{\textcolor{black}{#1}}
\newcommand{\Rev}[1]{\textcolor{black}{#1}}
\newcommand{\Rmin}[1]{\textcolor{black}{#1}}
\usepackage[numbers]{natbib}

\def\tsc#1{\csdef{#1}{\textsc{\lowercase{#1}}\xspace}}
\tsc{WGM}
\tsc{QE}
\tsc{EP}
\tsc{PMS}
\tsc{BEC}
\tsc{DE}

\begin{document}
\let\WriteBookmarks\relax
\def\floatpagepagefraction{1}
\def\textpagefraction{.001}

\shorttitle{Assisting Mission-Critical Traffic Flows with Active Queue Management in Industrial Internet of Things}

\shortauthors{Wang et~al.}

\title [mode = title]{Assisting Mission-Critical Traffic Flows with Active Queue Management in Industrial Internet of Things}                      




\author[aff1]{Shuo Wang}
\ead{shuowang@swin.edu.au}

\author[aff2]{Jonathan Kua*}
\ead{jonathan.kua@deakin.edu.au}

\author[aff1]{Jiong Jin}
\ead{jiongjin@swin.edu.au}

\author[aff1]{Yew Wee Wong}
\ead{yewweewong@swin.edu.au}

\author[aff1]{Prem Prakash Jayaraman}
\ead{pjayaraman@swin.edu.au}

\author[aff3]{Zhibo Pang}
\ead{zhibo.pang@pku.edu.cn}

\cortext[cor1]{Corresponding author}

\affiliation[aff1]{
  organization={School of Engineering, Swinburne University of Technology},
  city={Melbourne},
  state={VIC},
  postcode={3122},
  country={Australia}
}

\affiliation[aff2]{
  organization={School of Information Technology, Deakin University},
  city={Geelong},
  state={VIC},
  postcode={3220},
  country={Australia}
}

\affiliation[aff3]{
  organization={School of Advanced Manufacturing and Robotics, Peking University},
  city={Beijing},
  country={China}
}

\begin{abstract}
Mission-critical Industrial Internet of Things (IIoT) traffic flows require bounded network latency and jitter guarantees to ensure the safe functioning of critical industrial infrastructure. These flows are typically communicated via commodity network routers with conventional First-In-First-Out (FIFO) buffers. \JK{FIFO has proven to be the culprit of the well-known bufferbloat phenomenon, and the deployment of Active Queue Management (AQM) schemes have demonstrated significant performance improvements for latency-sensitive applications over the Internet in the IT domain. However, the bufferbloat phenomenon and the efficacy of AQM schemes have not been studied in IIoT-based OT domain. In this paper, we propose the use of AQM as a lightweight and non-intrusive mechanism for assisting mission-critical traffic flows in IIoT networks.} Our experimental results demonstrated that multi-queue AQM schemes provide substantial flow isolation and capacity sharing benefits, and significantly improve the performance of mission-critical traffic flows under network pressure. We further provide deployment recommendations based on our experimental insights.
\end{abstract}


\maketitle

\section{Introduction}

\Rev{The rapid technological advancements and innovations in Industrial Internet of Things (IIoT) and Industrial Cyber-Physical Systems (ICPS) have propelled industrial automation systems towards a fully-digitized and Artificial Intelligence (AI)-driven Industry 4.0 paradigm~\cite{wu2025embodied,xia2025emerging}. In recent years, research efforts from academic and industrial communities have focused on shifting factory automation systems towards a more modular Cloud-based architecture, such as the newly proposed ``Cloud-Fog Automation'' architecture~\cite{Cloud-Fog_Automation,jin2025cfa,mahmud2025trusted}. This architecture deploys lower layer applications such as supervisory controllers, machine controllers, I/O modules over wireless connections, whereas upper layers such as Enterprise Resource Planning (ERP) and Manufacturing Execution Systems (MES) are deployed in the Cloud.}

\JK{Cloud-Fog Automation represents an ambitious vision and major paradigm shift from current industrial architectures. Although Cloud-Fog Automation provides significant technical performance and CapEx/OpEx business benefits, it requires considerable amount of time and effort to be fully realized, primarily due to its disruptive nature to legacy systems.} Many businesses could not afford the substantial downtime and cost-inducing upgrades required for their current automation infrastructure. The return on investments from such business decisions and deployments may not be immediately evident. \R{However, businesses typically value and welcome non-intrusive technical decisions that have been experimentally-validated into their existing infrastructure~\cite{lezzi2018cybersecurity}.}


To this end, we propose the use of Active Queue Management (AQM) schemes in an IIoT-enabled industrial automation systems to drive the transition towards Cloud-Fog Automation. \Rmin{The deployment of AQM is a backward-compatible, lightweight, and non-intrusive mechanism for improving network performance in IIoT, as opposed to new deterministic networking technologies which incur extensive developmental and deployment costs. Typical industrial automation systems use commodity network routers to provide connectivity for a broad range of operational IIoT devices and applications. Network routers (with their proprietary vendor software or open-source software) typically (by default) use the conventional \textit{drop-tail} or First-In-First-Out (FIFO) queue management scheme due to its protocol and performance stability.}

\Rmin{However, FIFO has proven to be the culprit of the \textit{bufferbloat} phenomenon over the Internet (\textit{IT domain}), where unmanaged router buffers at either home gateways or Internet Service Providers (ISPs) lead to excessive delays and performance degradation in latency-sensitive and transactional traffic flows~\cite{Gettys:2011:Bufferbloat}. However, this phenomenon has not been studied in IIoT environments (\textit{OT domain}), partly due to the common perception that high-bandwidth local area network connectivity are able to support the relatively low-rate industrial control applications in IIoT.} The latency and reliability requirements of mission-critical applications (such as process control data, machine-to-machine communication, safety data, and control system signals, and so forth) is paramount. 

\Rev{\Rmin{Unlike conventional Internet environments, IIoT systems in the OT domain contain heterogeneous traffic flows that require not only throughput performance, but also predictable timing behavior, bounded latency and stable jitter characteristics.} Many industrial traffic flows (e.g., OPC UA telemetry, machine-control signaling, robotic coordination and supervisory process communication) operate at relatively low data rates yet remain highly sensitive to transient queue buildup and timing perturbations. In addition, modern industrial environments increasingly integrate hybrid wired/wireless connectivity, edge-cloud coordination and heterogeneous industrial subsystems through commodity IT-based networking infrastructure.} \Rmin{In such environments, unmanaged queueing behavior may introduce excessive delays and timing instability despite apparently sufficient bandwidth provisioning.}

\Rev{In traditional Internet applications, network congestion primarily degrades the Quality of Experience (QoE). However, queue-induced latency inflation in industrial environments may directly impact operational stability, synchronization and fail-safe system behavior. \Rmin{These characteristics motivate an experimental investigation into the behavior and deployment implications of modern AQM schemes in realistic IIoT environments.} Our work focuses on studying the use of state-of-the-art AQM schemes in IIoT, and investigating how they can be deployed to assist mission-critical flows and alleviate network performance bottlenecks (paralleling the benefits in the IT domain, as recommended in RFC7567~\cite{rfc7567}).}

\begin{figure*}[t]
\centering
\begin{subfigure}[t]{.45\textwidth}
    \includegraphics[scale=0.35]{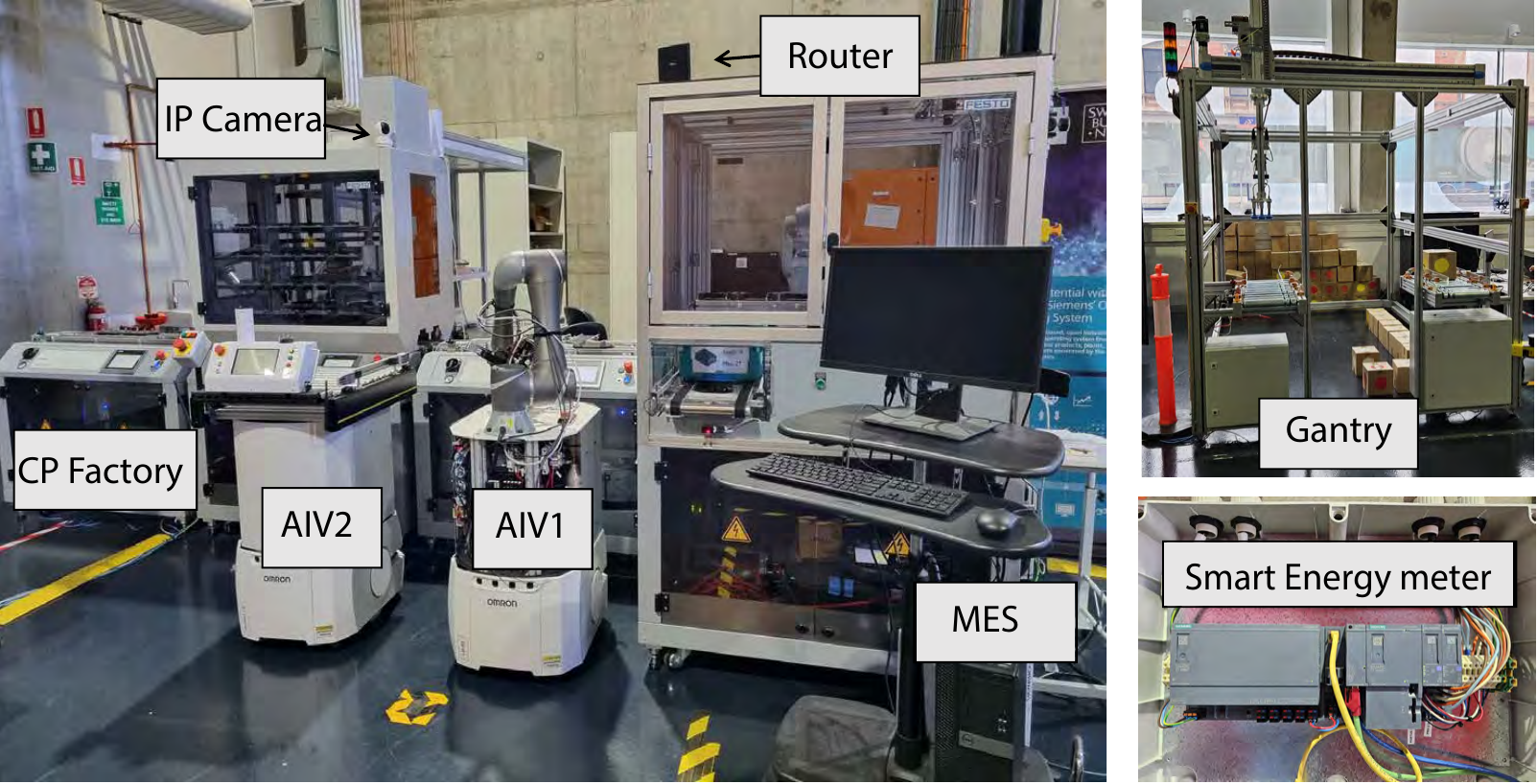}
    \centering
    \caption{\R{Physical experimental testbed setup}} \label{fig:FoFtestbed}
\end{subfigure}\hfill%
\begin{subfigure}[t]{.45\textwidth}
    \includegraphics[scale=0.26]{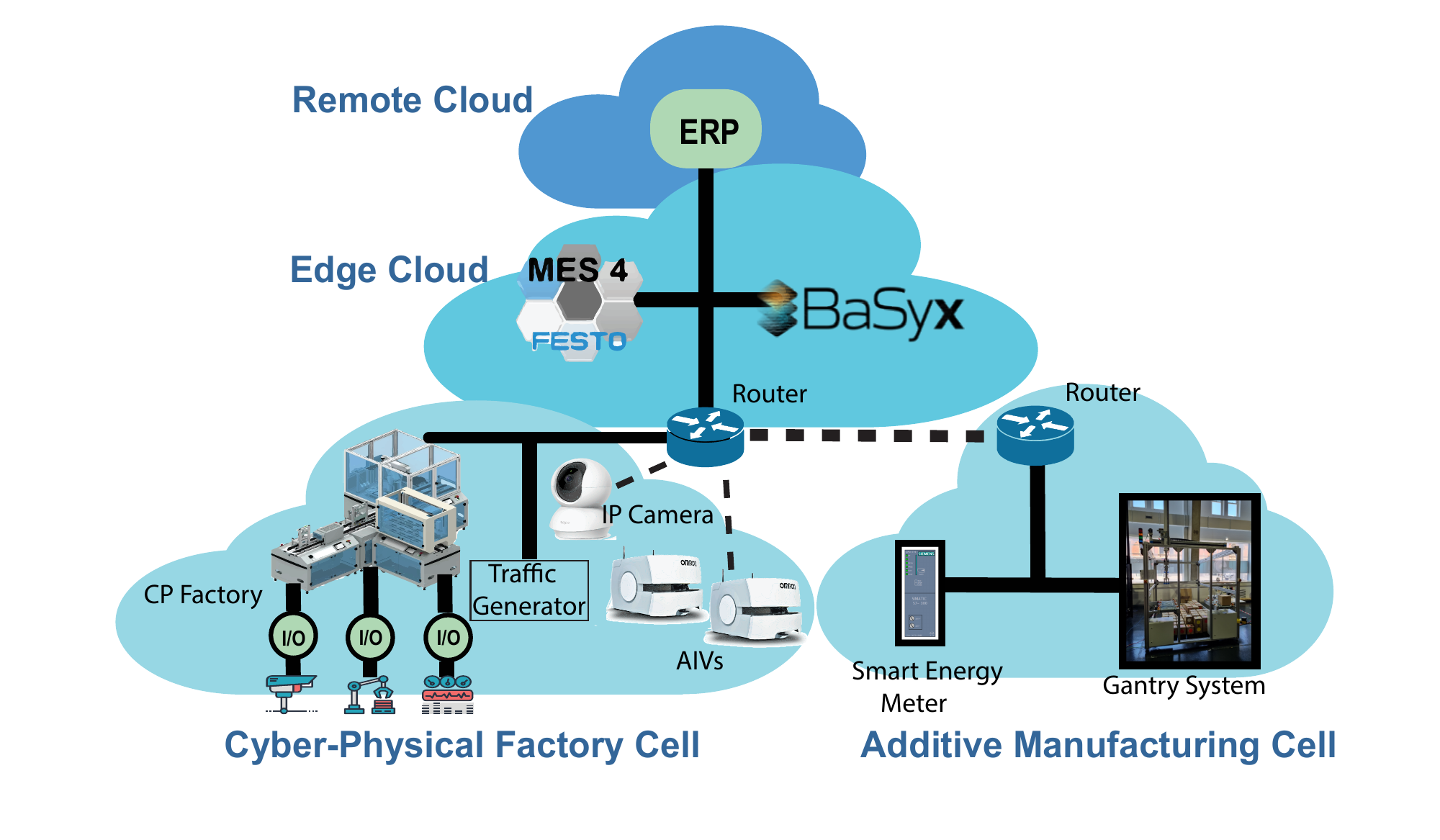}
    \centering
    \caption{\R{Network topology and configuration}} \label{fig:FoFtestbedtopology}
\end{subfigure}\hfill%
\caption{\Rmin{(a) Experimental testbed emulating (b) Cloud-Fog Automation in a hybrid wired/wireless multi-cell smart manufacturing factory. The Cyber-Physical Factory cell consists of two AIVs, a FESTO\textsuperscript{\tiny\textregistered} CP factory, IP camera and traffic generator, all connected to the MES via the router. The Additive Manufacturing cell consists of the Gantry system and Siemens smart meter, both connected to BaSyX via a wireless link.}}
\label{fig:FoFtestbedtopology_both}
\kern-0.8em
\end{figure*}

\Rmin{In this work, we emulate the shift towards Cloud-Fog Automation by constructing a sophisticated smart manufacturing experimental testbed using real-world industrial equipment with realistic hybrid multi-cell factory scenarios, and evaluated the performance of traffic flows generated by IIoT applications with or without the presence of AQM schemes. We compare the default FIFO queue management scheme with three representative AQM schemes: \textit{Controlled Delay} (CoDel, defined in RFC8289~\cite{rfc8289}, representing single-queue AQM), \textit{FlowQueue-CoDel} (FQ-CoDel, defined in RFC8290~\cite{rfc8290}, representing multi-queue AQM), \textit{Common Applications Kept Enhanced} (CAKE, representing set-associative-based AQM~\cite{cake_aqm}). Our experimental results demonstrated the significant flow isolation and capacity sharing benefits of deploying AQM in IIoT networks.}

\Rev{In summary, our work offers the first real-world experimental validation of how AQM schemes can assist mission-critical IIoT traffic flows. While existing AQM schemes are not new, their adoption in industrial settings has been limited. Hence, we provide experimental evidence that multi-queue AQM schemes significantly outperform FIFO and single-queue mechanisms in protecting latency-sensitive flows. Our key contributions are as follows:
}

\begin{enumerate}[(i)]
    \item \Rmin{Identify unmanaged queue buildup and bufferbloat as the cause of performance bottleneck in IIoT, which motivates the use of AQM as a lightweight, backward-compatible and non-intrusive mechanism for assisting mission-critical industrial traffic flows.}

    \item \Rev{Experimentally characterize the operational behavior of representative AQM schemes in a real-world smart manufacturing IIoT testbed comprising heterogeneous wired/wireless industrial subsystems, OPC UA-based communication, robotic coordination traffic, real-time monitoring applications and competing elastic traffic flows.}

    \item \Rmin{Analyze the impact of AQM schemes on mission-critical industrial applications' throughput stability, latency performance, jitter characteristics, flow isolation and operational responsiveness under congested and uncongested conditions.}

    \item \Rmin{Provide practical deployment recommendations and operational insights into the deployment of AQM in industrial automation systems based on our experimental observations and in-depth analysis from realistic industrial network scenarios.}
\end{enumerate}

The rest of the paper is organized as follows. Section~\ref{Section:Related_Work} provides background information and related work, Section~\ref{Section:Methodology} presents our experimental testbed and evaluation methodology, and \Rmin{Section~\ref{Section:Evaluation} discusses our experimental results and provides deployment recommendations.} Section~\ref{Section:Conclusions} concludes the paper and outlines potential future work.

\section{Background and Related Work}
\label{Section:Related_Work}

In this section, we provide background information on mission-critical IIoT traffic flows and AQM schemes, and discuss related work.

\subsection{\JK{Traffic management for mission-critical IIoT flows}}
\label{subsection:mission_critical}

\Rmin{In IIoT environments, mission-critical traffic refers to communication flows where timely and reliable delivery is necessary to maintain correct operational behavior, process coordination, safety functions or control responsiveness. Examples include machine-control signaling, robotic coordination traffic, supervisory process communication, OPC UA telemetry and safety-related monitoring data. Unlike conventional best-effort Internet traffic, these traffic flows are often highly sensitive to transient delay inflation, jitter variation and communication instability (even at relatively low data rates).}

\Rev{Latency guarantees extend beyond achieving low average delay in industrial environments. Rather, it emphasize bounded and predictable timing behavior. Many industrial applications rely on periodic communication cycles, synchronized operations and deterministic control responsiveness. Excessive delay variation or transient queue buildup may negatively affect operational stability, synchronization accuracy and fail-safe behavior (even when overall throughput remains sufficient).}

A crucial characteristic of IIoT is the need to ensure the timeliness of data transmission, with bounded latency and jitter guarantees, particularly for mission-critical applications. \JK{Control and data packets need to be transmitted and received within a strict deadline with minimum jitter (little variation in packet arrival). This represents `real-time' and `deterministic' network communication for mission-critical IIoT traffic flows.} The Industrial Internet Consortium (IIC) categorizes criticality into low, medium and high~\cite{industrial2019time}. Traffic types with \textit{high} criticality necessitate the fulfillment of their Quality of Service (QoS) guarantees (such as latency, jitter, and packet loss) satisfied to ensure a functioning system. If QoS guarantees are not met, it will lead to critical system malfunction. Applications with \textit{medium} criticality do not require their QoS standards to be consistently fulfilled; while unmet requirements may lead to performance degradation, it will not result in system failure. \JK{\textit{Low} criticality refers to applications that are not directly relevant to the critical part of the system. These applications tolerate packet losses and retransmissions, and do not require latency and jitter guarantees.}


\subsection{\JK{Standards and protocols for deterministic networking}}

\JK{Current approaches for ensuring determinism primarily rely on wired connections (e.g., Industrial Ethernet), modified protocols (e.g., MQTT~\cite{iotj_mqtt}, OPC UA~\cite{opc_ua}, EtherCAT~\cite{EtherCat}, PROFINET~\cite{Profinet}), network computing architectures~\cite{iotj_sdn_iiot} and integrated industrial networks (e.g., as proposed by the Internet Engineering Task Force (IETF)~\cite{draft-iotops-km-iiot-frwk-00,draft-tang-iiot-architecture-00}). Recent industrial developments include IETF's Deterministic Networking (DetNet) as defined in RFC8578~\cite{rfc8578,drl_detnet}, and IEEE 802.1's Time-Sensitive Networking (TSN)~\cite{iotj_tsn_routing_scheduling, iotj_tsn_software_defined}, which focuses on high precision traffic scheduling and prioritization to ensure determinism and end-to-end guarantees over wired and wireless links.}



\begin{table}
\centering
\caption{\JK{Queue Configurations (Baseline FIFO and AQM schemes)}}
\label{tab:AQM_configurations}
\begin{tblr}{|Q[c,1.5cm]|Q[c,6cm]|}
\hline
\textbf{\JK{Queue}} & \textbf{Configuration} \\
\hline 
\JK{FIFO (baseline, no AQM)} &  Queue size = 500 packets\\ 
\hline
\JK{CoDel AQM} & Queue size = 1000 packets, \Rmin{$T_{interval}$ = 100 ms} \Rmin{$T_{target}$ = 5 ms} \\ 
 \hline
\JK{FQ-CoDel AQM} & \shortstack{Queue size = 1000, \Rmin{$T_{interval}$ = 100 ms}, \\ Flows = 1024, Quantum = 1514 bytes} \\ 
\hline
\JK{CAKE AQM} &  \shortstack{RTT scheme = LAN, Diffserv mode = 3, \\ Flow mode = triple isolate}  \\ 
\hline
\end{tblr}
\end{table}

\subsection{\JK{State-of-the-art AQM schemes}}


\Rev{Conventional FIFO queueing passively buffers packets until congestion becomes severe. AQM schemes are designed to proactively regulate/manage queue occupancy to reduce excessive delay buildup and improve responsiveness under congestion (i.e., minimizing overall system latency and jitter). Single-queue approaches such as CoDel primarily control queue delay within a shared bottleneck queue, whereas multi-queue mechanisms such as FQ-CoDel and CAKE additionally provide flow isolation and fairer capacity sharing among competing traffic flows. \Rmin{These characteristics are particularly relevant in industrial environments where heterogeneous latency-sensitive and elastic traffic co-exist over shared communication infrastructure.} AQM mechanisms dynamically drop or mark packets before the queue becomes full, hence signaling congestion to the endpoints\footnote{\JK{In our work, FIFO is used strictly as a baseline for comparison and is not considered a form of AQM.}}.
}

\JK{PIE~\cite{rfc8034} estimates queuing delay from buffer occupancy and egress rate. It admits packets freely for the first 150ms of an empty queue, then probabilistically drops packets based on a periodically adjusted drop probability (every 30ms by default), targeting a delay of \Rmin{15 ms}. When the drop probability is below 10\%, ECN-enabled flows are marked instead of dropped. PIE bases its decisions on both the current queuing delay and its trend. CoDel~\cite{rfc8289} tracks the minimum queuing delay over fixed intervals (initially \Rmin{100 ms}) using packet timestamps. If the minimum delay exceeds a target value throughout an interval, CoDel begins dropping packets based on a control law and continues until the delay falls below the target. ECN marking is also used when supported.}

\JK{FQ-CoDel~\cite{rfc8290} classifies traffic flows into one of 1024 sub-queues by hashing the flow's five-tuple. CoDel is used to manage each sub-queue independently, and the modified Deficit Round Robin (DRR) scheduler is used to dequeue a `quantum' of bytes (1500 bytes by default), allowing FQ-CoDel to prioritize `sparse' (latency-sensitive) traffic flows. It provides fair sharing of network resources and bandwidth sharing while controlling the flows' queueing delay.}

\JK{CAKE~\cite{cake_aqm} is a shaping capable AQM that uses a combination of COBALT (CoDel+BLUE AQM), a traffic shaper, and a variant of DRR++ scheduler for flow isolation. It uses an 8-way associative hashing to eliminate hash collisions for placing flow packets into the independently-created sub queues, with each sub-queue running its own AQM state, and prioritizes latency-sensitive flows. CAKE provides several extra options, such as priority queuing with a simplified DiffServ implementation, overhead compensation, and Round Trip Time (RTT) tuning.}

\subsection{\JK{Related work and research gaps}}
\JK{Existing studies in the literature focused on evaluating and improving AQM schemes in consumer and commercial broadband networks in the wider Internet~\cite{grazia2022aggregating,kua2017using,iotj_drl_aqm_sensor_networks}. The work in~\cite{morato2021network} studied both the TCP incast and congested bottleneck problems in IIoT using network simulators. Other recent studies have focused on deep mathematical formulation and reasoning for designing cooperative TCP and AQM systems in the Internet~\cite{abdelmoniem2023enhancing}.}

\Rev{New deterministic networking standards such as TSN are gaining traction in the community. TSN standards outline the data scheduling, synchronization and traffic shaping mechanism to achieve the specified latency requirements. TSN operates directly at Layer 2 (data link/MAC layer) and bypasses the TCP/IP layer~\cite{wollschlaeger2017future,tran2025empowering}.} \Rmin{It requires dedicated TSN switches (and infrastructure) that implements TSN stream scheduling/prioritization mechanisms to guarantee the deterministic latency requirements.} In contrast to TSN, Active Queue Management (AQM) reduces intra-network latency at the network and transport layers without requiring specialized switches or infrastructure. AQM is supported by most commodity routers via the operating system kernel, and it manages internal buffer queues by creating logical sub-queues. It monitors queuing delays and proactively drops or marks packets to restrain aggressive flows and prioritize latency-sensitive traffic. While TSN offers deterministic latency and reliability, its adoption is limited by high deployment costs and poor interoperability with existing protocols such as EtherCAT and PROFINET. We present a low-cost, backward-compatible solution through the use of AQM in current IIoT infrastructure to enhance the performance of mission-critical traffic across both wired and wireless links.


\Rev{Recent research efforts in industrial networking have increasingly focused on latency-sensitive communication~\cite{cheng2018industrial}, deterministic networking and QoS provisioning for IIoT/cyber-physical systems. Emerging approaches include TSN-based deterministic communication~\cite{jiang2023assessing}, edge-assisted industrial networking architectures, OPC UA real-time extensions~\cite{lee2025opc,ladegourdie2022performance,martins2023cnc} and adaptive traffic-management mechanisms for heterogeneous industrial workloads~\cite{kumar2023analysis}. Modern AQM schemes such as CoDel, FQ-CoDel and CAKE have demonstrated significant latency-reduction benefits in broadband Internet environments by mitigating excessive queue buildup and improving flow isolation under network congestion. However, research on the operational impact of these queue management mechanisms in realistic OT-domain/IIoT environments remain limited. In particular, there are limited studies on the impact of AQM schemes on heterogeneous mission-critical traffic flows operating across hybrid wired/wireless industrial communication infrastructure, which require stringent operational timing and tight coupling with cyber-physical processes~\cite{hawaou2024industry}. On the other hand, existing IIoT networking studies typically assume sufficiently provisioned local connectivity, rely on deterministic networking mechanisms such as TSN, or evaluate traffic behavior primarily through simulation-based models.}

\subsection{\JK{Novelty and contributions}}

\JK{In summary, most existing studies on IIoT networking assume high-bandwidth local connectivity or rely on deterministic mechanisms (e.g., TSN, EtherCAT) without accounting for the effects of unmanaged queue buildup. As a result, the bufferbloat phenomenon (extensively studied in the IT domain) has received limited attention in the OT domain. For instance, the aforementioned prior works either use simulation-based models or pre-configured deterministic settings that do not reveal the queuing delay inflation caused by FIFO~\cite{grazia2022aggregating,kua2017using,iotj_drl_aqm_sensor_networks,morato2021network}. In contrast, our work highlights this performance bottleneck through real-world experiments and evaluates how AQM schemes can alleviate it across hybrid wired/wireless IIoT environments. This fills an important gap in the literature by showing that even in seemingly over-provisioned industrial networks, queuing effects can severely degrade mission-critical traffic unless explicitly managed.}

\Rmin{To the best of our knowledge, this work represents the \textit{first in-depth real-world experimental study} on investigating and evaluating the use of AQM schemes in the OT-domain/IIoT environments.} It is our hope that this foundational work can motivate industrial network researchers and practitioners to adopt AQM in IIoT and advance IIoT-specific AQM innovations.

\begin{table*}[t]
\centering
\caption{Traffic Flows' Characteristics and Functionalities}
\label{tab:flows_characteristics}
\begin{tblr}{|Q[c,3cm]|Q[c,2cm]|Q[c,1.5cm]|Q[p,10cm]|}

\hline
\textbf{Name} & \textbf{Application-layer Bitrate} & \textbf{Criticality} & \textbf{Traffic Properties and Functionalities}\\ 
\hline 
Autonomous Intelligent Vehicle (AIV) 1
& \Rmin{9.3 Kbps}, WiFi & High & Periodic, real-time monitoring of the movement and location with robot arm operation control, taking 3D printed sensor back housing to scan cell for quality inspection.
\\
\hline 
Autonomous Intelligent Vehicle (AIV) 2
& \Rmin{9 Kbps}, WiFi & High  & Periodic, real-time monitoring of movement and location, exchanging PCB boxes for assembly robot to assist with part assembly.\\ 
\hline 
IP Camera &  \Rmin{1 Mbps}, WiFi & Medium & Cyclic, real-time IP-based video streaming for surveillance, monitoring workplace environment and detecting intrusions.\\ 
\hline 
\R{FESTO\textsuperscript{\tiny\textregistered} Cyber-Physical (CP) Factory} & \R{\Rmin{0.5 Kbps}, Ethernet} & \R{Low}  & \R{Sporadic, alarm and events data from CP Factory, modular production line testbed and programmed for the assembly of sensor products.}\\ 
\hline 
\R{Gantry and BaSyX\textsuperscript{\tiny\textregistered}} &  \R{\Rmin{2 Kbps}, Ethernet and WiFi} & \R{High}  & \R{OPC UA flow from Omron PLC to BaSyx, which is a Digital Twin implementation for real-time control and monitoring of Gantry's operation.}   \\
\hline 
\R{Siemens\textsuperscript{\tiny\textregistered} Smart Meter} &  \R{\Rmin{25 Kbps}, Ethernet and WiFi} & \R{High}  & \R{OPC UA over TCP traffic flow for real-time energy monitoring of the 3D printing process.}  \\
\hline
\R{Background Traffic} &  \R{Elastic rate, Ethernet} & \R{Low}  & \R{Elastic (\textit{i.e.,} consuming all available bandwidth) TCP CUBIC traffic flows, representing background traffic with high-bandwidth utilization in congested networks.} \\
\hline 
\end{tblr}
\kern-1.2em
\end{table*}

\section{Experimental Methodology}
\label{Section:Methodology}

\R{In this section, we present our experimental testbed setup and evaluation methodology for studying the benefits and impact of FIFO and AQM on mission-critical IIoT flows.}

\Rev{We note that this work does not propose a new AQM scheme or modify the internal mechanisms of existing queue management schemes. The primary objective of our work is to experimentally characterize and compare the operational behavior of representative state-of-the-art AQM schemes under realistic industrial IIoT environments. The evaluated AQM schemes are selected to represent distinct queue management design approaches and operational characteristics. CoDel represents single-queue delay-based AQM, FQ-CoDel combines queue-delay management with per-flow isolation and fair scheduling, while CAKE extends flow-isolated queue management with integrated shaping and enhanced traffic-handling mechanisms. These schemes are widely deployed (enabled by default in many ISPs' network routers/home gateways) and representative of modern low-latency queue management approaches supported by commodity networking infrastructure. \Rmin{The evaluated network conditions are designed to emulate realistic industrial communication scenarios in which IIoT traffic flows co-exist with competing elastic traffic over shared communication infrastructure.}}

\subsection{\JK{System design and architecture}}

\JK{The primary goal of our experimental setup is to emulate a realistic Cloud-Fog Automation network in which mission-critical traffic coexists with background or elastic traffic flows. A central edge router is used to reflect common industrial deployment patterns where traffic from multiple sub-systems converges through a bottleneck queue before reaching upstream services (e.g., cloud dashboards, digital twins). This router becomes the enforcement point for queue management policies.}

\JK{To evaluate AQM under typical network load conditions, we introduce a background traffic generator that simulates bandwidth-hungry flows such as firmware updates, data logging, or external system backups. This configuration allows us to analyze how different queue management schemes (FIFO, CoDel, FQ-CoDel, CAKE) affect the latency and reliability of time-sensitive IIoT communications.}

\subsection{\JK{Experimental testbed setup and implementation}}
\label{label:experimental_testbed_setup}
\JK{We built a real-world experimental testbed in our University's smart manufacturing testlab to investigate the benefits and impact of AQM on mission-critical traffic flows in IIoT environments. Figure~\ref{fig:FoFtestbedtopology_both} illustrates our physical experimental testbed setup and the corresponding network topology/configuration. Technical details and specifications of each key component are described in the following.}

\R{\textit{Routers.} We use the MikroTik\textsuperscript{\tiny\textregistered} hAP ac2 as our bottleneck router which connects the Cyber-Physical Factory cell to the Edge Cloud. An additional MikroTik router is connected to the bottleneck router via a wireless link to support the Additive Manufacturing cell, which is located in a farther physical location. Both internal routing software run Linux-based Router OS V7 that implements FIFO and CoDel, FQ-CoDel, CAKE AQM schemes\footnote{\JK{\url{https://help.mikrotik.com/docs/display/ROS/Queues}}}, and supports traffic shaping with Hierarchical Token Bucket (HTB) to provide a configurable bottleneck between the MES/BaSyX at the Edge Cloud level and field equipment/end-devices. }

\JK{\textit{Background traffic generator.}
To emulate background industrial traffic and stress-test AQM's impact on evaluated IIoT traffic flows, we use \textit{iperf3} to generate elastic TCP flows that maximize available bandwidth usage. The underlying TCP algorithm runs TCP CUBIC, which is known for its effective bandwidth utilization, and represents the default TCP algorithm of most industrial operating systems.}

\textit{FESTO\textsuperscript{\tiny\textregistered} Cyber-Physical (CP) Factory\footnote{\JK{\url{https://www.festo.com/au/en/e/technical-education/training-concepts/highlights/training-factories/cp-systems-all-round-i4-0-training-factories-id_32122/}}} and Manufacturing Execution System (MES).} CP Factory incorporates a suite of industrial functions into a modular and expandable factory. It consists of standard transportation modules for linear conveyor belts and branch modules. Our testbed features mobile modules with electrical, pneumatic, and mechanical interfaces, allowing for easy topological adjustments, utilizing Siemens\textsuperscript{\tiny\textregistered} SIMATIC ET200SP PLC modules. Additionally, a symbiotic cell extended from the CP Factory enables the concurrent operation of collaborative robots (Cobots) and operators. Our MES operates on an Industrial PC (IPC), which serves as the central configuration, monitoring, and execution point for all machines and devices within the production cell. All control and process data to and from the MES are communicated via the MikroTik router.

\textit{\R{Omron\textsuperscript{\tiny\textregistered} Autonomous Intelligent Vehicles (AIVs).}} \JK{In our system, AIVs are deployed to assist human operators in performing quality inspection tasks on 3D-printed components. Two Omron LD-90 AIVs are utilized: AIV1 is equipped with a TM5-900 collaborative robotic arm and a PLC NX102-9020, while AIV2 operates without an onboard manipulator. The data traffic associated with AIV1 is categorized as mission-critical, as it encompasses time-sensitive control, coordination, localization, and situational awareness data streams between the robotic system and the MES. All communication is routed through the AIVs and a centralized bottleneck router. The job sequence proceeds as follows: upon receiving a task from the MES, AIV1 autonomously navigates to the designated workstation, retrieves the target components, and transports them to the 3D scanning cell. Once scanning is complete, a human operator performs a final verification, after which AIV1 returns to its docking location. Continuous and timely data exchange is essential to maintain operational integrity and avoid disruptions in the end-to-end production workflow.}

\R{\textit{Gantry and BaSyX\textsuperscript{\tiny\textregistered} systems.} A key component of our Additive Manufacturing cell is the Gantry and BaSyX system. Gantry is a crane-like system with industrial robots that operate on a mechanical framework that uses a mobile trolley over a linear bridge. Our setup uses Gantry as a smart inventory system for storing raw materials for 3D printing. It consists of a step motor-driven robot arm with a vacuum tip for picking and placing boxes and an Omron PLC NX102-9020 (from the AIVs) for logic control and data transmission. At the Edge Cloud level (in parallel with the MES), BaSyX\textsuperscript{\tiny\textregistered}\footnote{\url{https://eclipse.dev/basyx/}} is a Digital Twin middleware that we implemented for monitoring and controlling Gantry's operations. The traffic flows generated by Gantry are considered mission-critical as real-time inventory information is required to control and coordinate the 3D printing process.}

\R{\textit{Smart energy meter}. We use Siemens\textsuperscript{\tiny\textregistered} ET 200SP Analog input module AI energy meter 480VAC ST as our smart meter. It serves as the monitoring and reporting unit for the operation of the 3D printers in the Additive Manufacturing cell. Traffic flows generated by the smart meter to the BaSyX system are classified as mission-critical since the real-time energy consumption information of the printing process are required to optimize and improve the quality control of the 3D-printed parts while reducing material and energy wastage.}

\Rev{In the Additive Manufacturing cell, the PLCs driving the Gantry/BaSyx gantry subsystem and the smart-energy monitoring subsystem are linked via Ethernet to a local wireless access point. Both PLCs communicate using OPC UA. Process variables are published in real time, enabling client applications to subscribe to live telemetry streams. A UA-Expert OPC UA client continuously records the instantaneous energy consumption of the fused-deposition-modeling (FDM) printer. Although these OPC UA flows have a low data rate, their timeliness is critical; therefore, the proposed AQM schemes must prioritize and protect them.}


\begin{figure*}[t]%
\centering
\begin{subfigure}{.3\textwidth}
    \includegraphics[width=\linewidth]{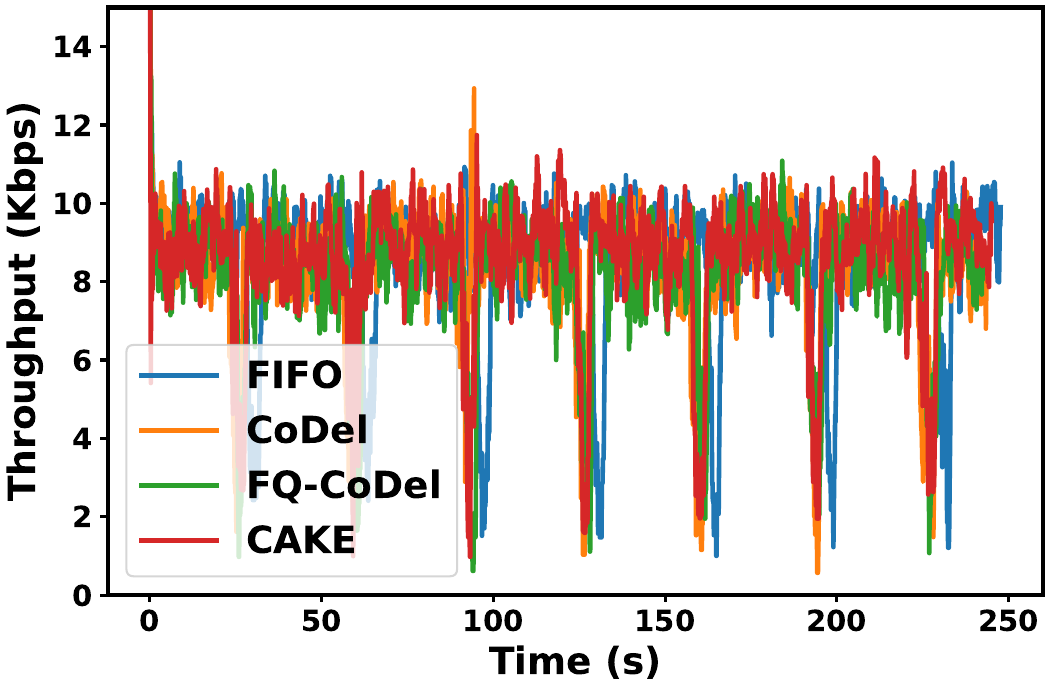}
    \caption{\Rmin{Throughput vs time (uncongested)}} \label{fig:AIVtimeseriesTP1}
\end{subfigure}\hfill%
\begin{subfigure}{.3\textwidth}
    \includegraphics[width=\linewidth]{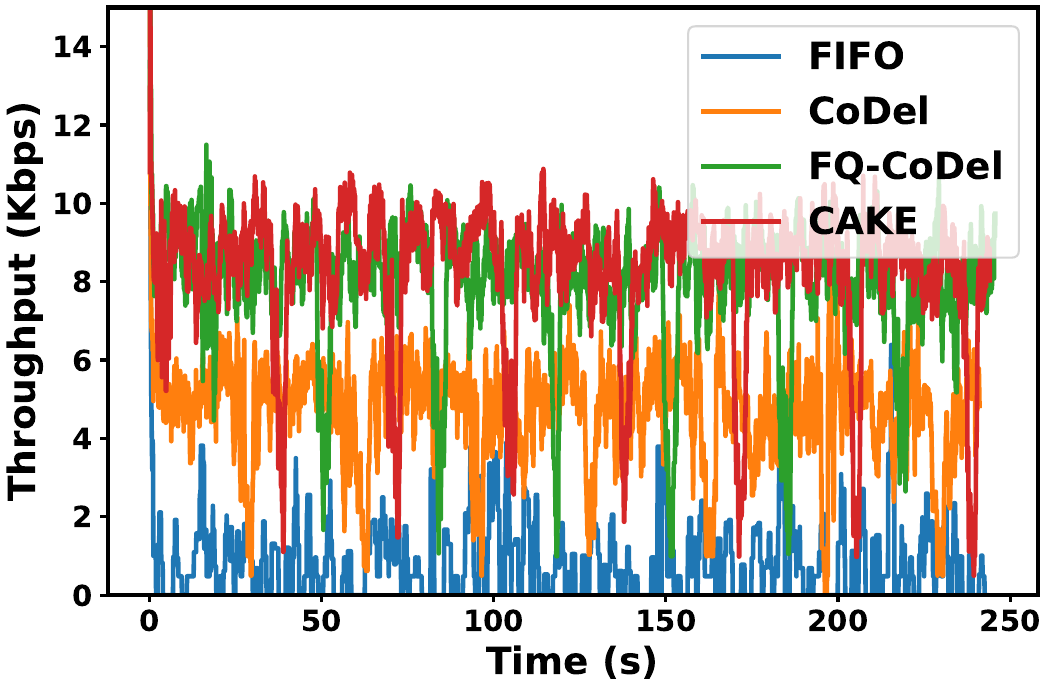}
    \caption{\Rmin{Throughput vs time (congested)}} \label{fig:AIVtimeseriesTP2}
\end{subfigure}\hfill%
\begin{subfigure}{.3\textwidth}
    \includegraphics[width=\linewidth]{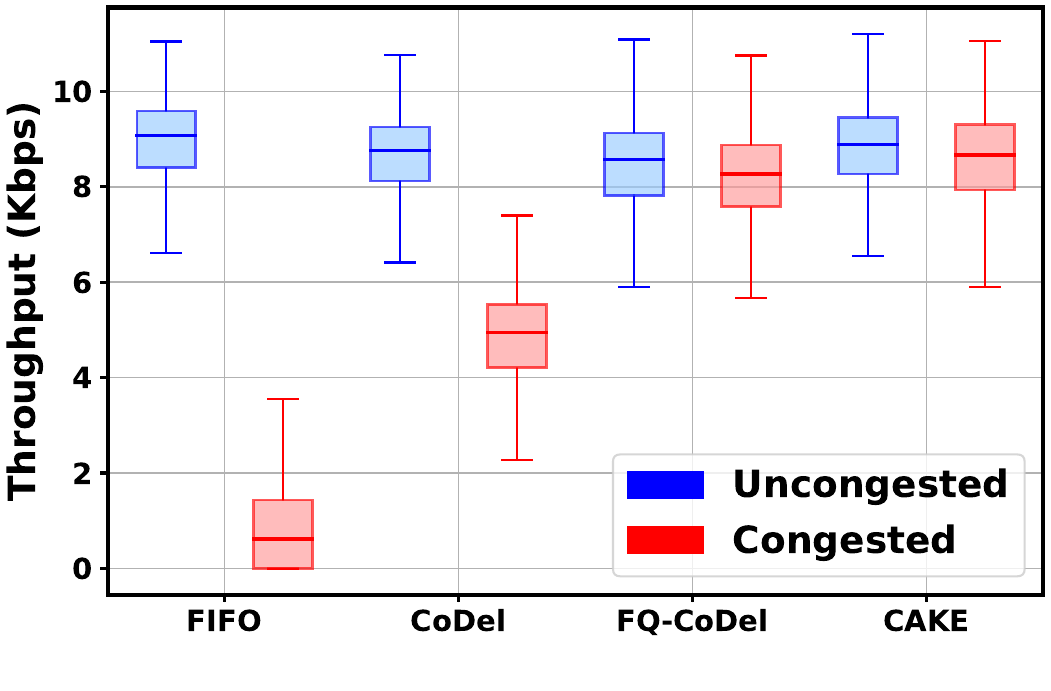}
    \caption{\Rmin{Boxplot of throughput}} \label{fig:AIVflowThroughputBOX}
\end{subfigure}\hfill%
\begin{subfigure}{.3\textwidth}
    \includegraphics[width=\linewidth]{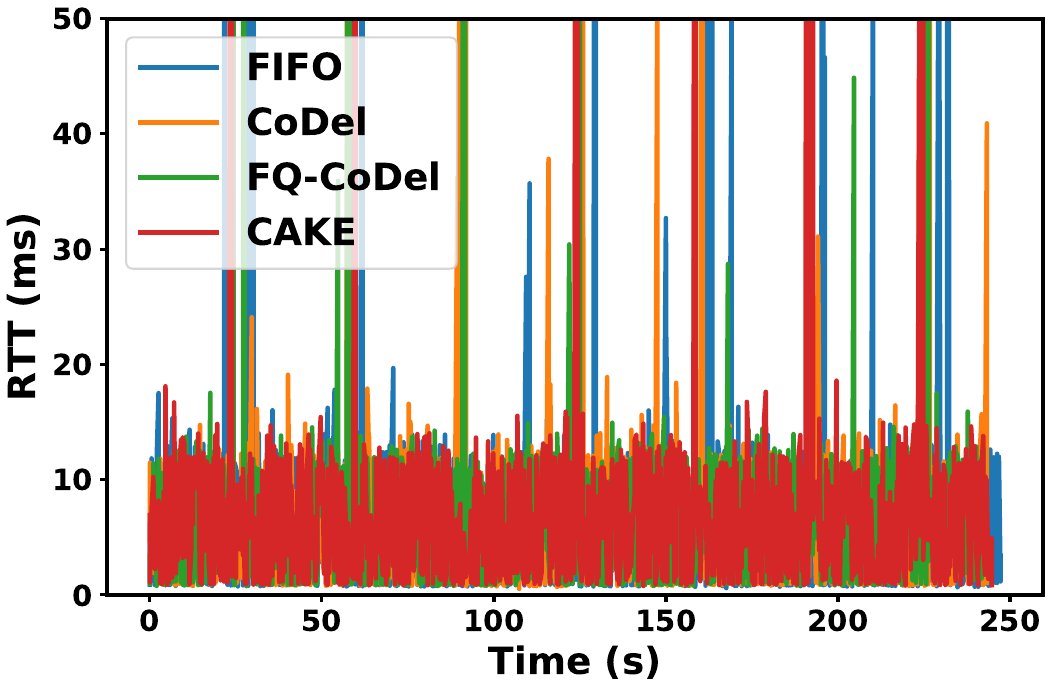}
    \caption{\Rmin{RTT vs time (uncongested)}} \label{fig:AIVtimeseriesRTT1}
\end{subfigure}\hfill%
\begin{subfigure}{.3\textwidth}
    \includegraphics[width=\linewidth]{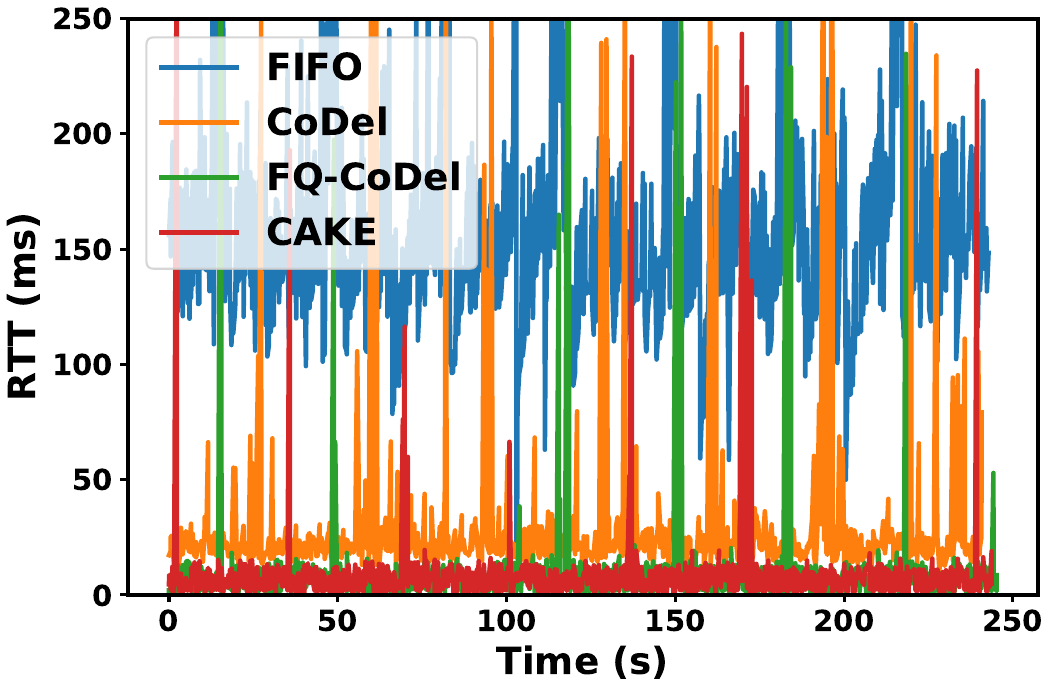}
    \caption{\Rmin{RTT vs time (congested)}} \label{fig:AIVtimeseriesRTT2}
\end{subfigure}\hfill%
\begin{subfigure}{.3\textwidth}
    \includegraphics[width=\linewidth]{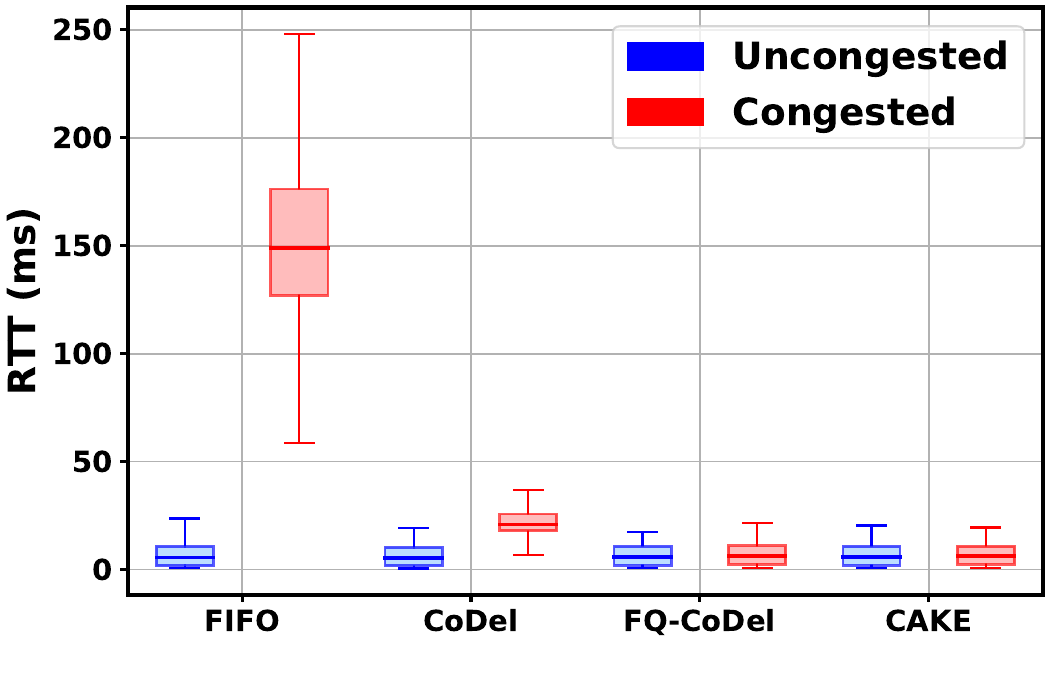}
    \caption{\Rmin{Boxplot of RTT}} \label{fig:AIVRTTBOX}
\end{subfigure}\hfill%
\caption{\Rmin{Time series and boxplots of throughput and RTT performance of AIV1's traffic flow in uncongested (a, d) and congested (b, e) network conditions in Scenario I.}}

 \label{fig:overall_results1}
\end{figure*}

\subsection{Performance metrics and network conditions}

\Rev{\Rmin{We measure and analyze the flows' throughput, RTT, jitter and packet-size distributions to characterize the operational communication behavior of the industrial traffic flows.} Full-length network packets are captured using \textit{tshark} at the MES' and BaSyX's Network Interface Card (NIC), for deep packet inspection and traffic analysis. Throughput is calculated with a sliding window to determine the number of bits received per second (bps) per traffic flow. RTT and delay measurements are derived from the MES and BaSyX's perspective (systems at the Edge Cloud level perspective). It is important to note that our RTT measurements differs from the connotation of RTT in conventional data networks. The RTT measurements presented in this work includes path delays, induced queuing delays and the computational time of the tasks. Jitter is calculated as the variation of packet arrival times and interpreted in the context of operational timing stability for mission-critical industrial applications, and the packet size distribution is analyzed accordingly.}

\JK{We emulate a Cloud-Fog Automation paradigm (Fig.~\ref{fig:FoFtestbedtopology}) where the MES is situated in the Edge Cloud, and data to/from the MES traverses through a bottleneck path in the operational factory environment. We set the bottleneck bandwidth at a symmetrical \Rmin{10 Mbps} for both uplink and downlink. The queue size for FIFO is configured at 500 packets (as per the Bandwidth Delay Product rule), and 1000 packets for CoDel, FQ-CoDel, and CAKE AQM (as per IETF's recommendations). Explicit Congestion Notification (ECN) is disabled at the end-hosts and router.} 

\Rev{Table~\ref{tab:AQM_configurations} summarizes the queue configurations and Table~\ref{tab:flows_characteristics} presents the flow characteristics of the diverse traffic types in our test environment. The queue management parameters are selected based on commonly recommended operational configurations from the corresponding IETF RFCs and vendor-supported implementations. FIFO queue sizing followed the Bandwidth Delay Product (BDP) guideline, while CoDel and FQ-CoDel parameters use their default operational interval and target settings recommended in RFC8289 and RFC8290. CAKE is configured using its LAN RTT profile and triple-isolate flow mode to reflect a representative low-latency industrial deployment scenario. }

\Rev{All experiments are conducted using fixed queue management configurations and consistent traffic-generation settings across the evaluated scenarios to ensure comparability of the observed operational behavior among FIFO, CoDel, FQ-CoDel and CAKE. Note that the objective of these settings is not parameter optimization, but rather an evaluation of representative real-world deployment behavior under realistic industrial traffic conditions.
}

\section{Performance Evaluation and Analysis}
\label{Section:Evaluation}

\Rev{In this section, we present our experimental findings and analysis of three representative industrial automation network scenarios. We then discuss deployment considerations and recommendations of AQM in industrial automation systems. The presented experimental scenarios are intentionally designed to represent an IIoT environment where heterogeneous mission-critical and background traffic flows co-exist over shared wired/wireless commodity networking infrastructure. The evaluated deployment reflects a transitional Cloud-Fog Automation setting in which industrial devices, PLCs, robotic systems, monitoring applications and Edge-Cloud services communicate through conventional IP-based routing infrastructure rather than a fully deterministic networking architecture. Our work focused on examining the scalability and operational robustness of AQM schemes under diverse industrial communication scenarios.}

\begin{figure*}[t]%
\centering
\begin{subfigure}{.25\textwidth}
    \includegraphics[width=\linewidth]{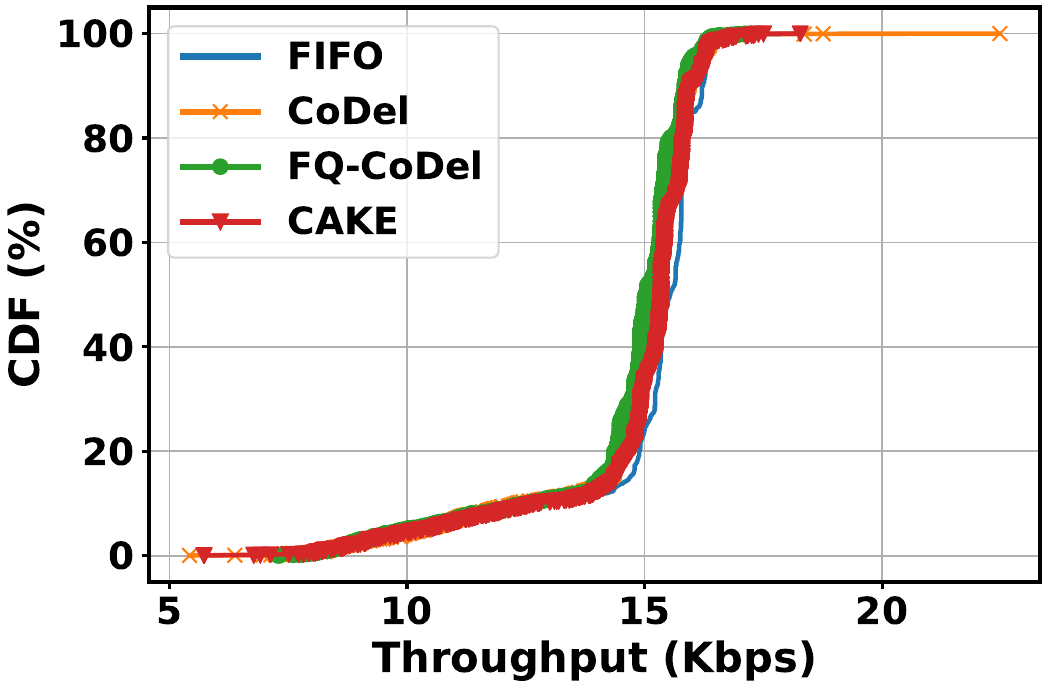}
    \caption{\Rmin{Throughput (uncongested)}} \label{fig:CDFAIVthroughput1}
\end{subfigure}\hfill%
\begin{subfigure}{.25\textwidth}
    \includegraphics[width=\linewidth]{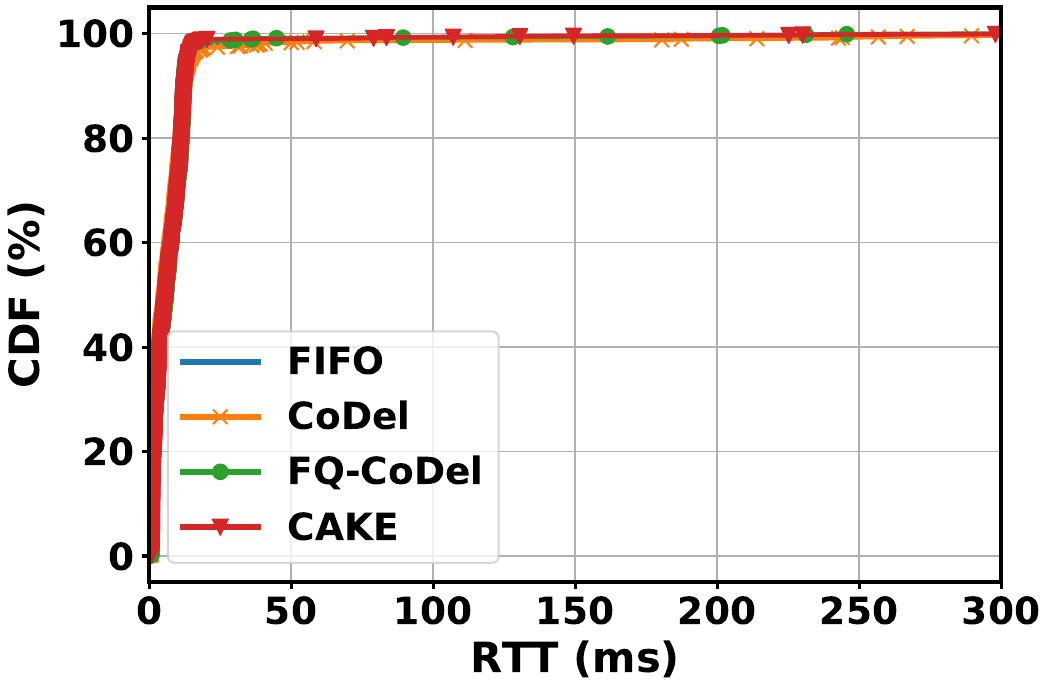}
    \caption{\Rmin{RTT (uncongested)}} \label{fig:CDFAIVrtt1}
\end{subfigure}\hfill%
\begin{subfigure}{.25\textwidth}
    \includegraphics[width=\linewidth]{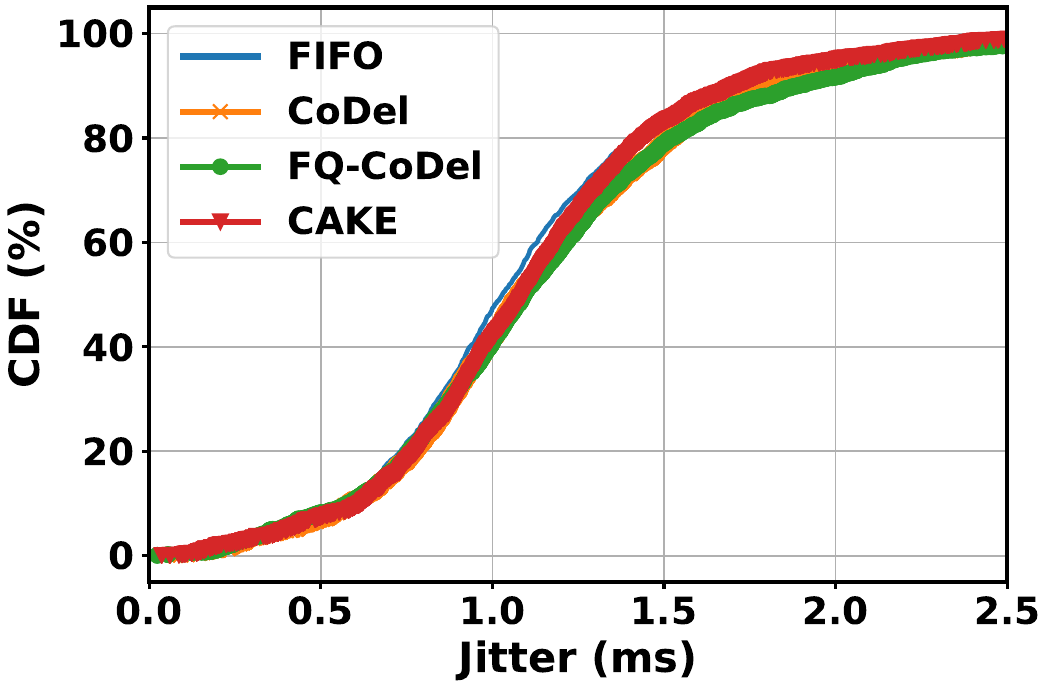}
    \caption{\Rmin{Jitter (uncongested)}} \label{fig:CDFAIVjitter1}
\end{subfigure}\hfill%
\begin{subfigure}{.25\textwidth}
    \includegraphics[width=\linewidth]{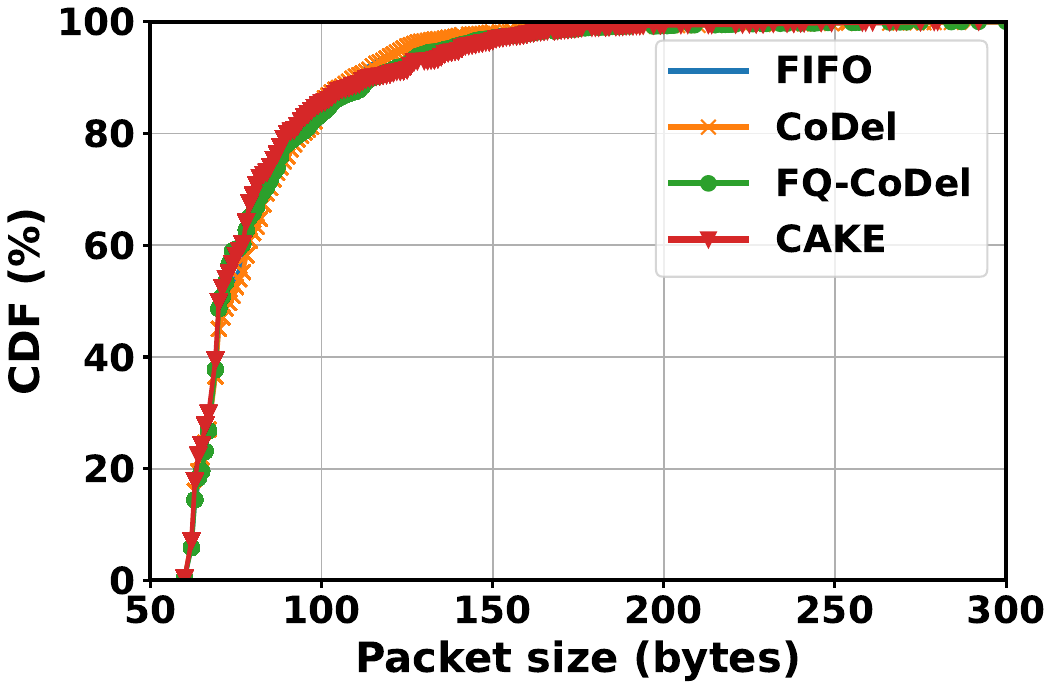}
    \caption{\Rmin{Packet size (uncongested)}} \label{fig:CDFAIVpacketdistr1}
\end{subfigure}\hfill%
\begin{subfigure}{.25\textwidth}
    \includegraphics[width=\linewidth]{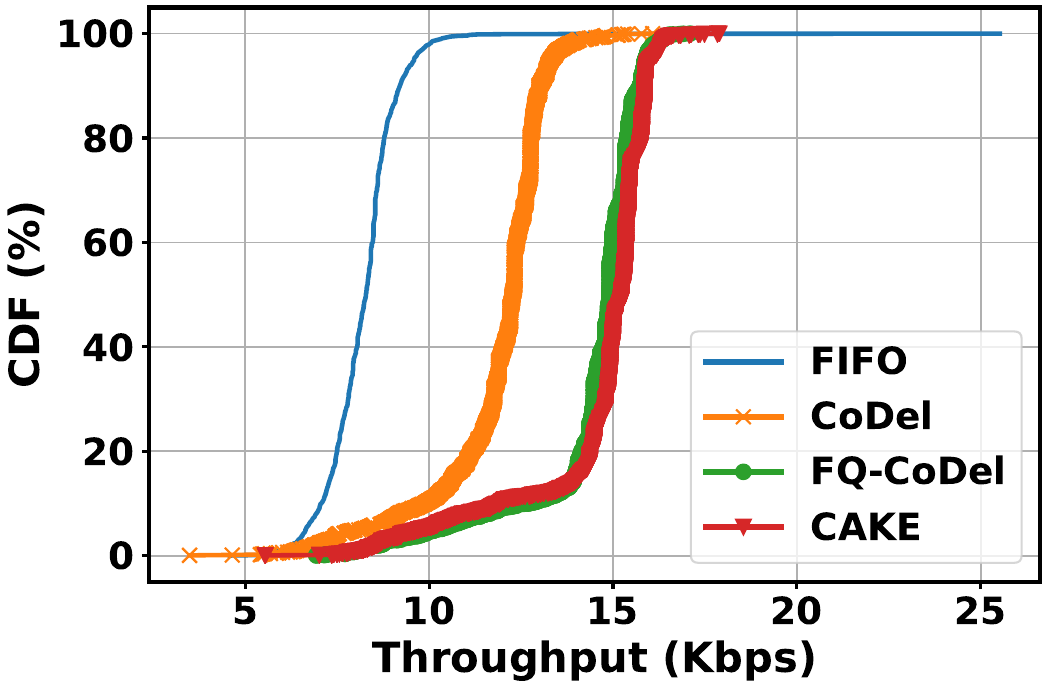}
    \caption{\Rmin{Throughput (congested)}} \label{fig:CDFAIVthroughput2}
\end{subfigure}\hfill%
\begin{subfigure}{.25\textwidth}
    \includegraphics[width=\linewidth]{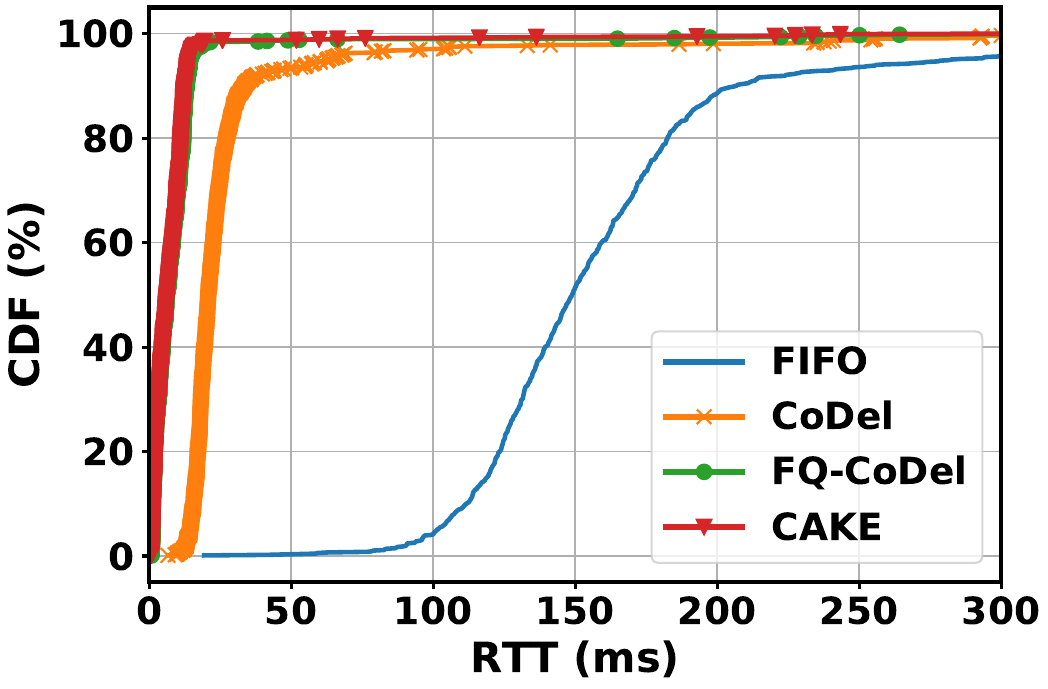}
    \caption{\Rmin{RTT (congested)}} \label{fig:CDFAIVrtt2}
\end{subfigure}\hfill%
\begin{subfigure}{.25\textwidth}
    \includegraphics[width=\linewidth]{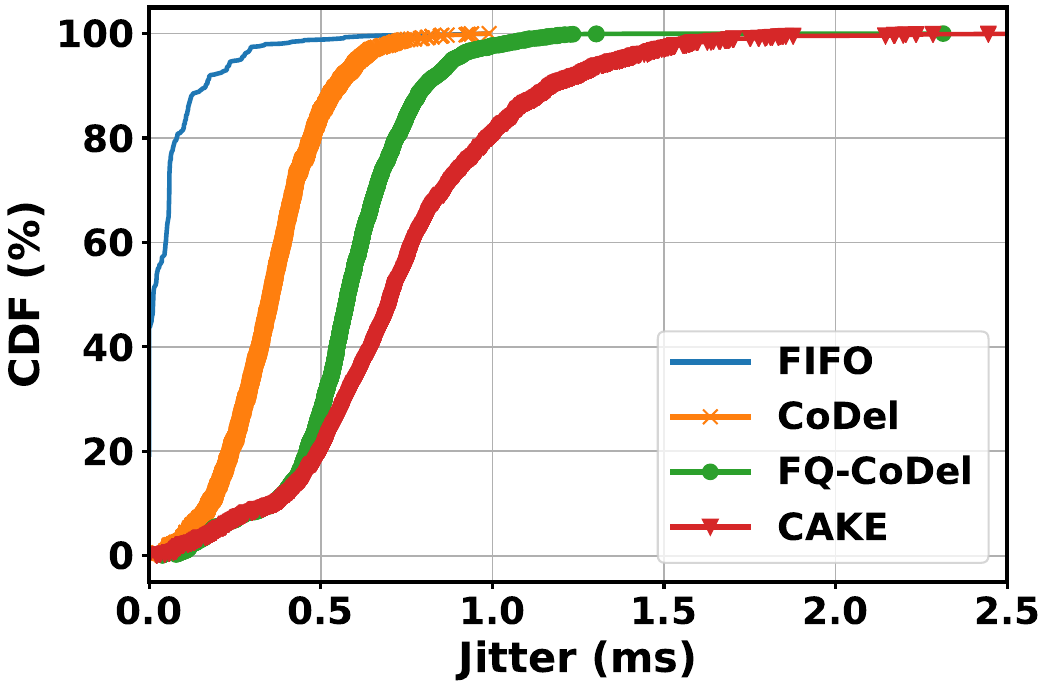}
    \caption{\Rmin{Jitter (congested)}} \label{fig:CDFAIVjitter2}
\end{subfigure}\hfill%
\begin{subfigure}{.25\textwidth}
    \includegraphics[width=\linewidth]{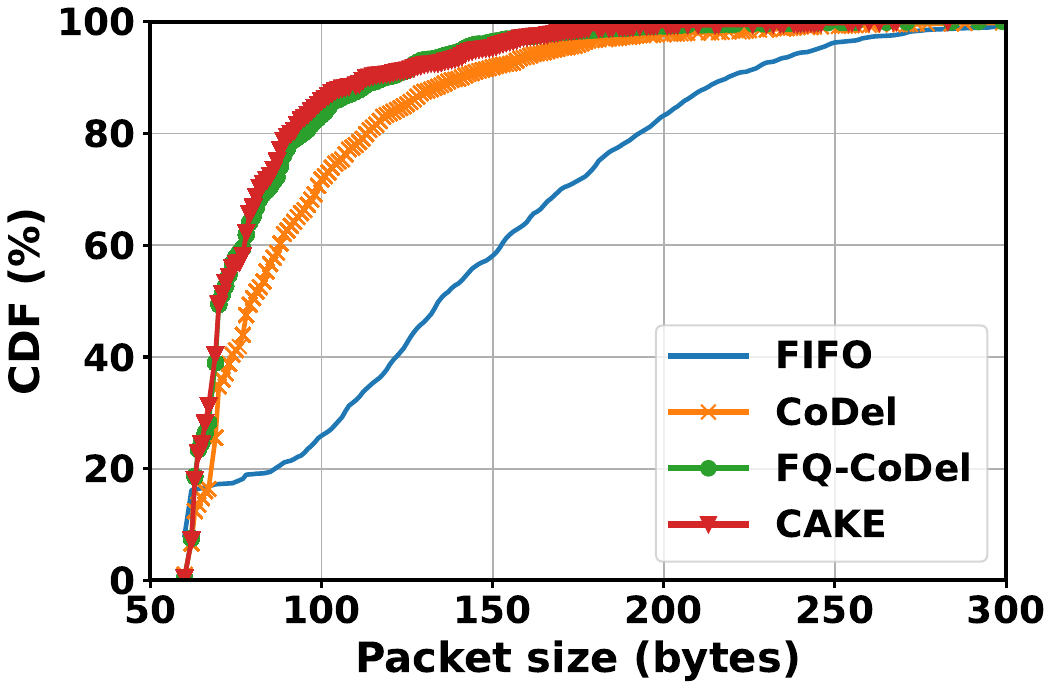}
    \caption{\Rmin{Packet size (congested)}} \label{fig:CDFAIVpacketdistr2}
\end{subfigure}\hfill%
\caption{\Rmin{CDF plots of throughput, RTT, jitter and packet size distribution of AIV1's traffic flow in uncongested (a)-(d) and congested (e)-(h) network conditions in Scenario I.}} \label{fig:CDF1}
\end{figure*}

\subsection{Scenario I: Typical industrial automation network}
\Rmin{Scenario I focuses on how different queue disciplines impact mission-critical IIoT flows (AIV1 flow over wireless) in a typical industrial automation network (within the Cyber-Physical Factory cell as illustrated in Fig.~\ref{fig:FoFtestbedtopology}), under congested/uncongested environments (with/without elastic TCP (\textit{iperf}) flows). In this scenario, the field devices/equipment (comprising of two AIVs: AIV1 and AIV2, FESTO CP Factory and \textit{iperf} traffic generator) are connected to the MES IPC via the router. The performance of AIV1 under uncongested environment over FIFO bottleneck represents the baseline and ground truth for the IIoT flows. We examine how AQM (either benefit or impede) the performance of these flows under congested/uncongested conditions.}


\subsubsection{The impact on throughput}
Fig.~\ref{fig:AIVtimeseriesTP1} and~\ref{fig:AIVtimeseriesTP2} present AIV1's throughput across time (250 secs) in both uncongested and congested scenarios. As a ground truth, AIV1 generates \Rmin{15 Kbps} in an uncongested environment across all queue disciplines, i.e., FIFO, CoDel, FQ-CoDel and CAKE. This performance indicates that AQM has no negative impact on AIV1's flow, which provides us with a baseline. The impact of FQ-CoDel and CAKE is similar, where AIV1's throughput maintains the same \Rmin{15 Kbps} in FQ-CoDel and CAKE, while the throughput drops to \Rmin{5 Kbps} in CoDel. This performance is expected, as FQ-CoDel and CAKE both implements hashing and multiple sub-queue functions for even-capacity sharing, and protect latency-sensitive flows (which is reflected in AIV1's flow in our experiments), whereas CoDel manages the packets within a single queue. The conventional FIFO queue shows the expected bufferbloat phenomenon, with a drastic 80\% drop in throughput to \Rmin{1 Kbps}, which renders AIV1/robotic arm unusable. \Rmin{Fig.~\ref{fig:AIVflowThroughputBOX},~\ref{fig:CDFAIVthroughput1} and~\ref{fig:CDFAIVthroughput2} compare the throughput performance across queue disciplines under congested/uncongested conditions}. The benefit of deploying AQM (specifically FQ-CoDel and CAKE) for preserving AIV1's flow throughput, and by extension, all latency-sensitive industrial mission critical flows is pronounced.

\begin{figure*}[t]%
\centering
\begin{subfigure}[t]{.3\textwidth}
    \includegraphics[width=\linewidth]{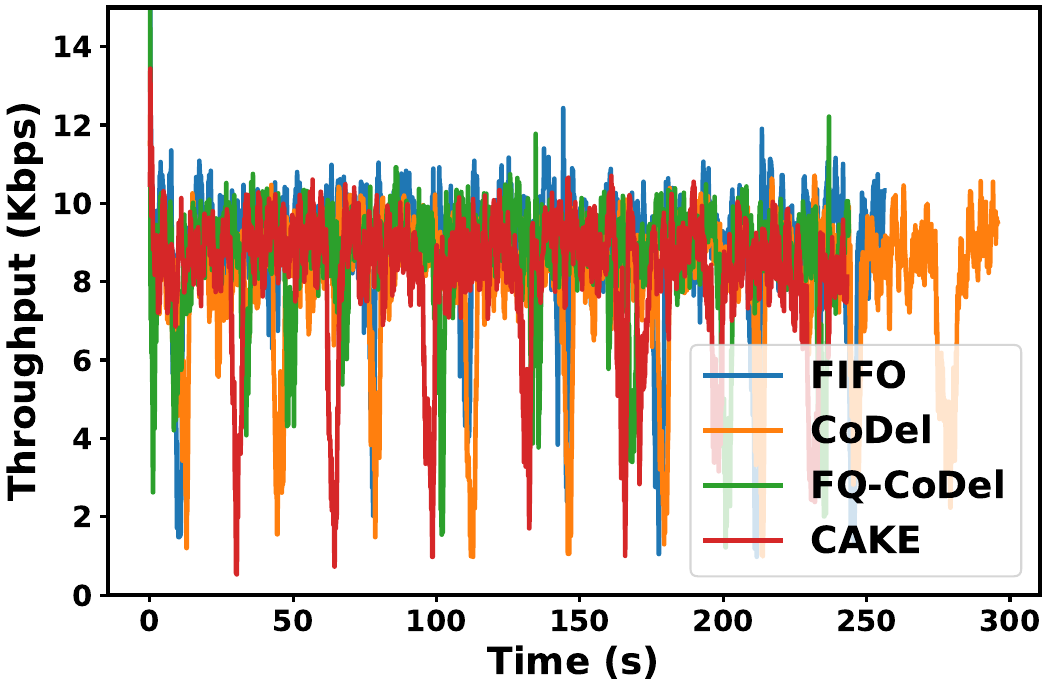}
    \caption{\Rmin{Throughput vs time (uncongested)}} \label{fig:AIVtimeseriesTP3}
\end{subfigure}\hfill%
\begin{subfigure}[t]{.3\textwidth}
    \includegraphics[width=\linewidth]{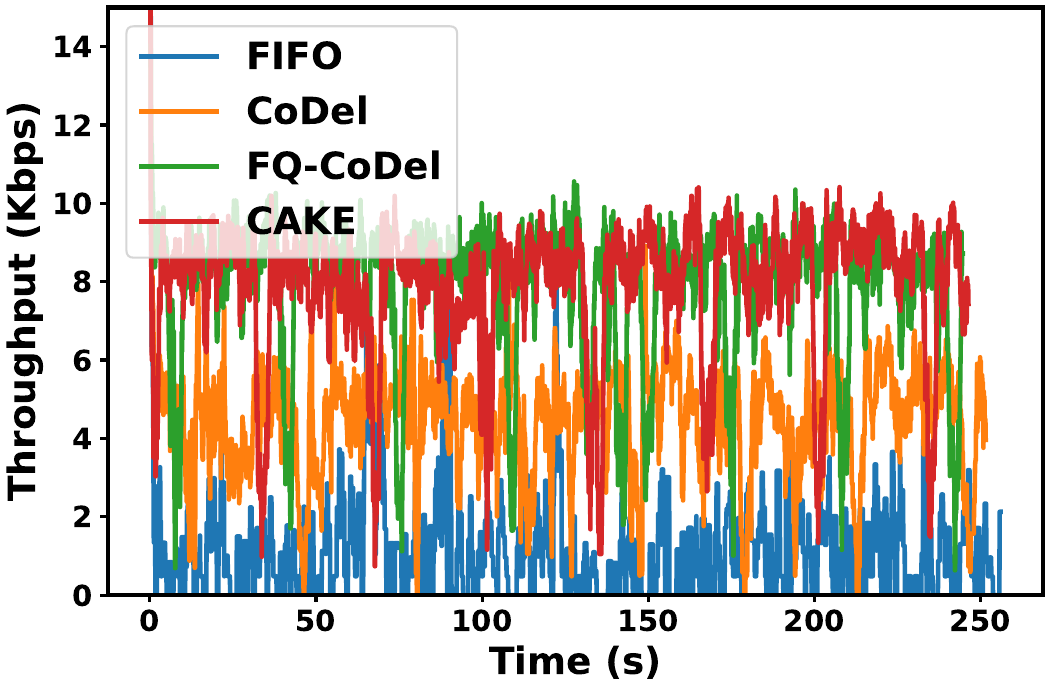}
    \caption{\Rmin{Throughput vs time (congested)}} \label{fig:AIVtimeseriesTP4}
\end{subfigure}\hfill%
\begin{subfigure}[t]{.3\textwidth}
    \includegraphics[width=\linewidth]{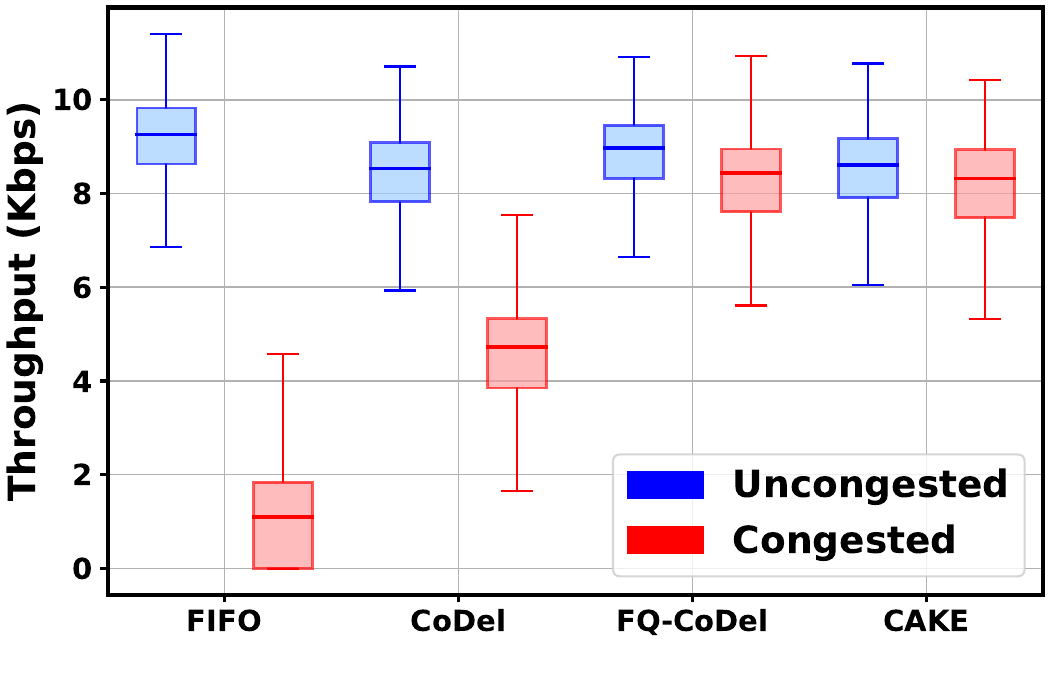}
    \caption{\Rmin{Boxplot of throughput}} \label{fig:AIVflowtpBOX2}
\end{subfigure}\hfill%
\begin{subfigure}{.3\textwidth}
    \includegraphics[width=\linewidth]{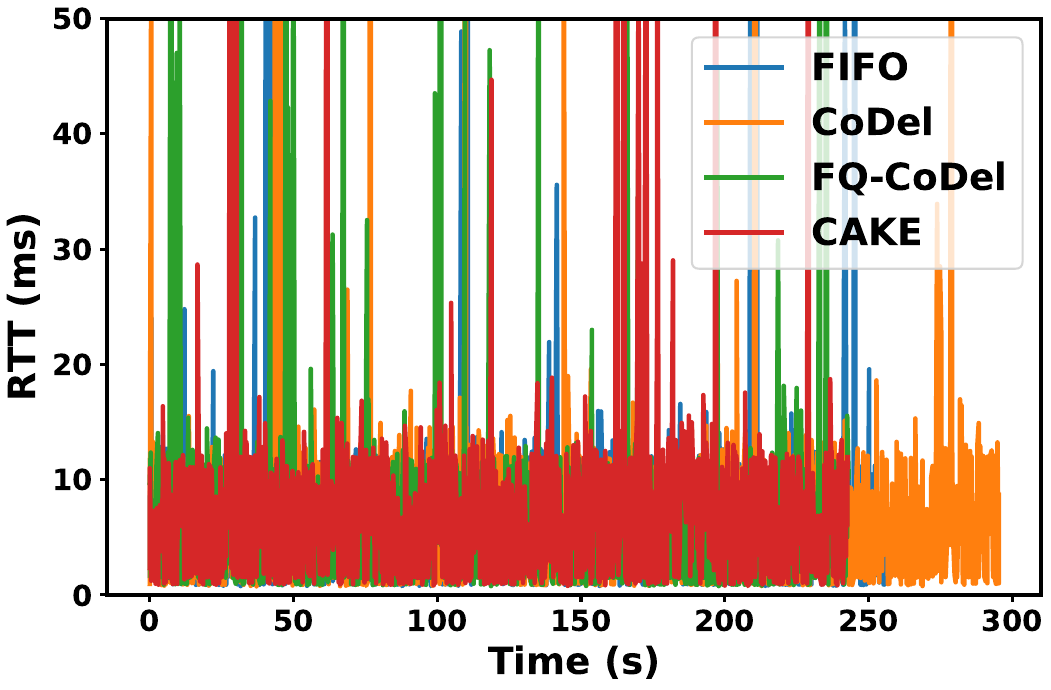}
    \caption{\Rmin{RTT vs time (uncongested)}} \label{fig:AIVtimeseriesRTT3}
\end{subfigure}\hfill%
\begin{subfigure}{.3\textwidth}
    \includegraphics[width=\linewidth]{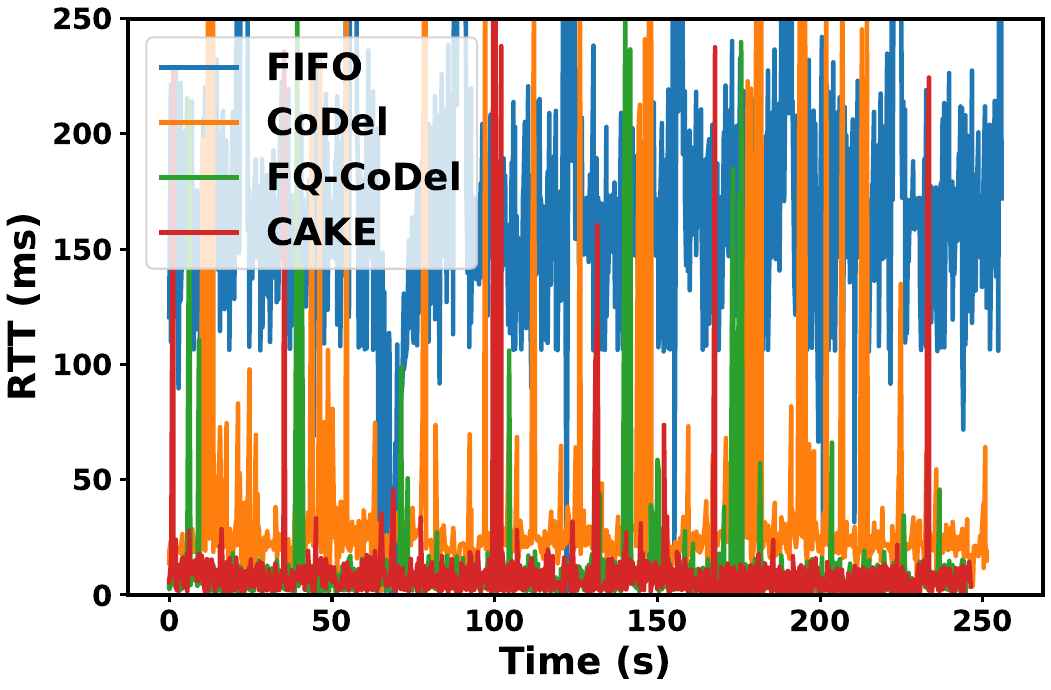}
    \caption{\Rmin{RTT vs time (congested)}} \label{fig:AIVtimeseriesRTT4}
\end{subfigure}\hfill%
\begin{subfigure}{.3\textwidth}
    \includegraphics[width=\linewidth]{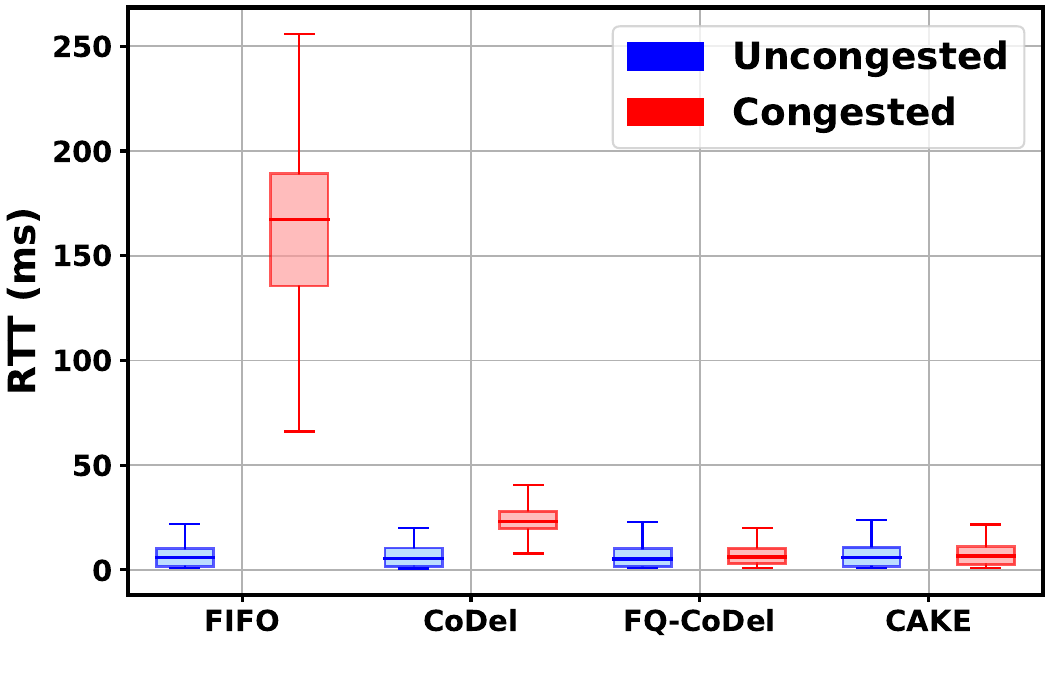}
    \caption{\Rmin{Boxplot of RTT}} \label{fig:AIVRTTBOX2}
\end{subfigure}\hfill%

\caption{\Rmin{Time series and boxplots of throughput and RTT performance of AIV1's traffic flow in uncongested (a, d) and congested (b, e) network conditions in Scenario II.}}

\label{fig:overall_results2}
\end{figure*}
\begin{figure*}[t]%
\centering
\begin{subfigure}{.25\textwidth}
    \includegraphics[width=\linewidth]{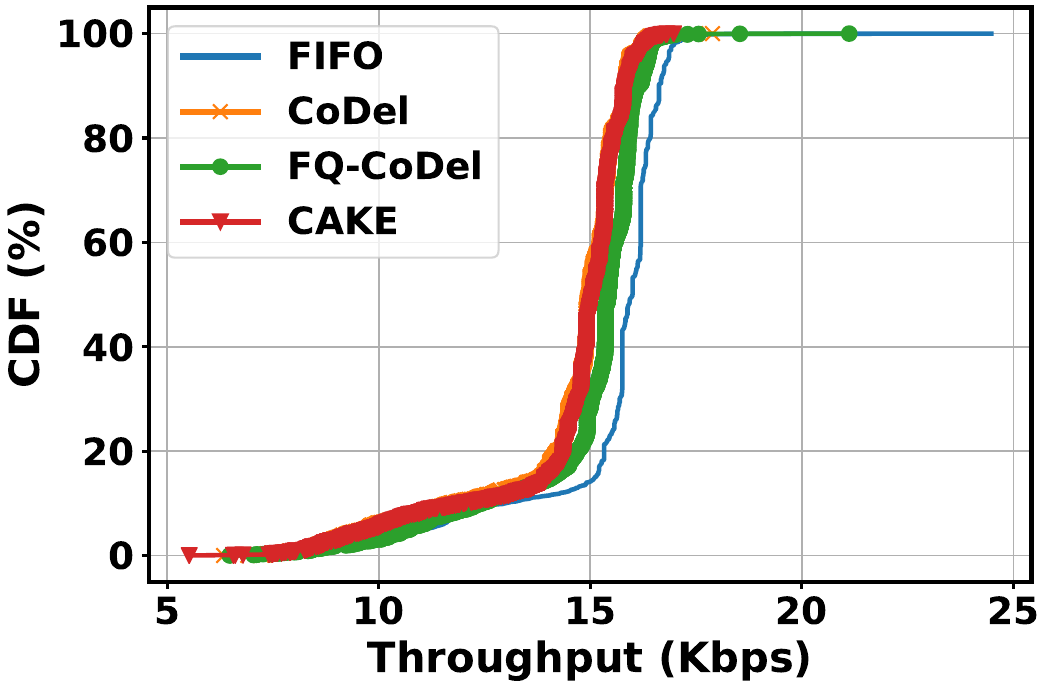}
    \caption{\Rmin{Throughput (uncongested)}} \label{fig:CDFAIVthroughput3}
\end{subfigure}\hfill%
\begin{subfigure}{.25\textwidth}
    \includegraphics[width=\linewidth]{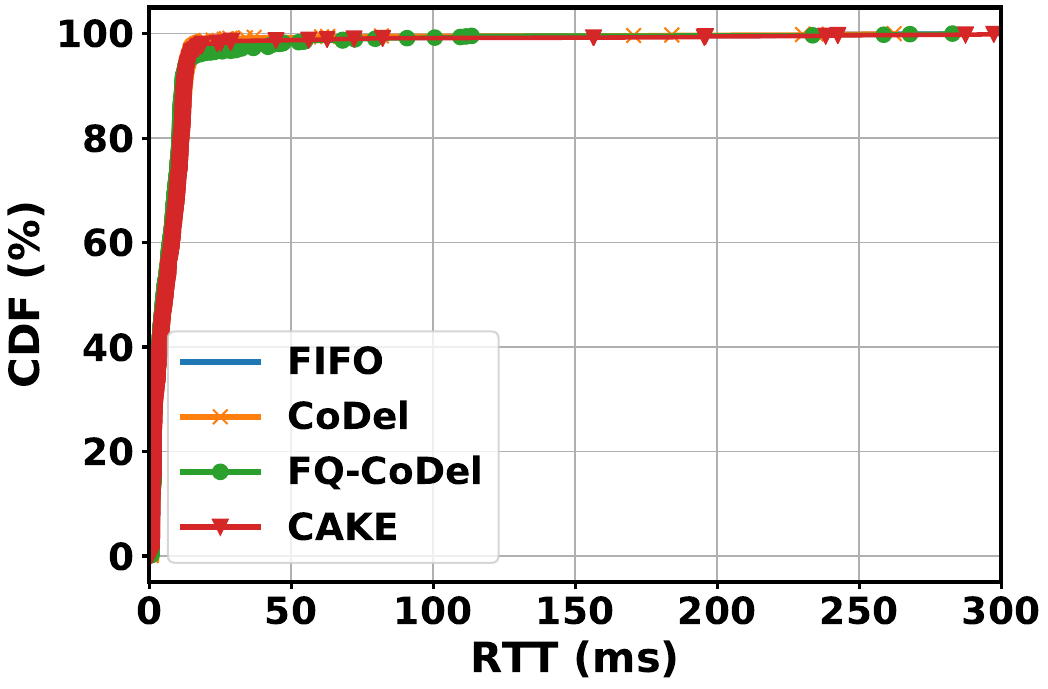}
    \caption{\Rmin{RTT (uncongested)}} \label{fig:CDFAIVrtt3}
\end{subfigure}\hfill%
\begin{subfigure}{.25\textwidth}
    \includegraphics[width=\linewidth]{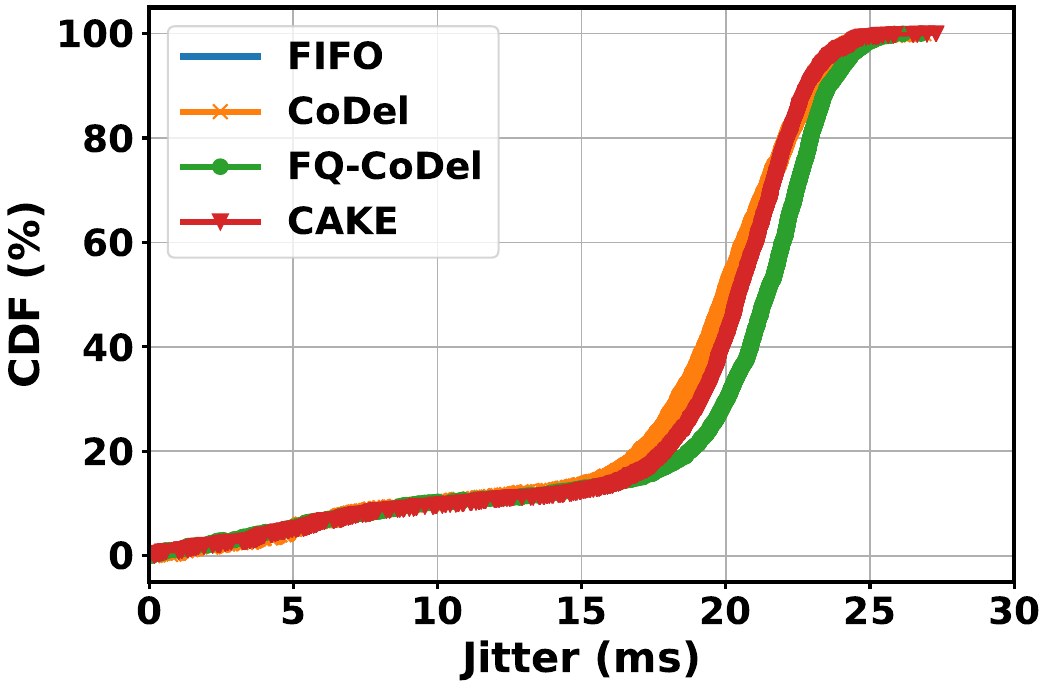}
    \caption{\Rmin{Jitter (uncongested)}} \label{fig:CDFAIVjitter3}
\end{subfigure}\hfill%
\begin{subfigure}{.25\textwidth}
    \includegraphics[width=\linewidth]{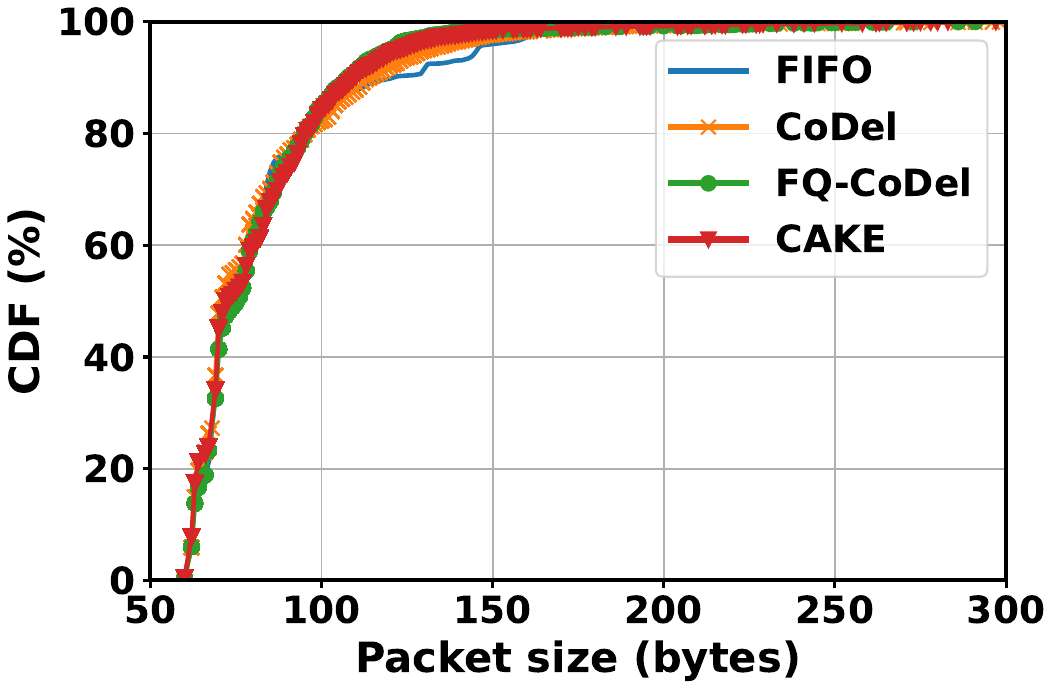}
    \caption{\Rmin{Packet size (uncongested)}} \label{fig:CDFAIVpacketdistr3}
\end{subfigure}\hfill%
\begin{subfigure}{.25\textwidth}
    \includegraphics[width=\linewidth]{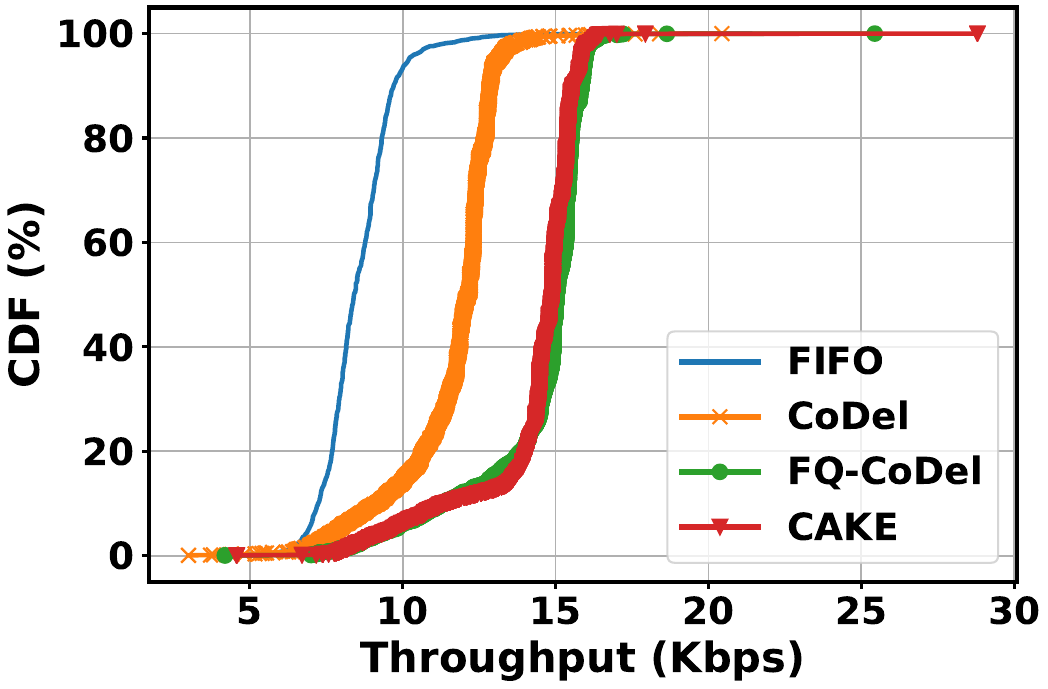}
    \caption{\Rmin{Throughput (congested)}} \label{fig:CDFAIVthroughput4}
\end{subfigure}\hfill%
\begin{subfigure}{.25\textwidth}
    \includegraphics[width=\linewidth]{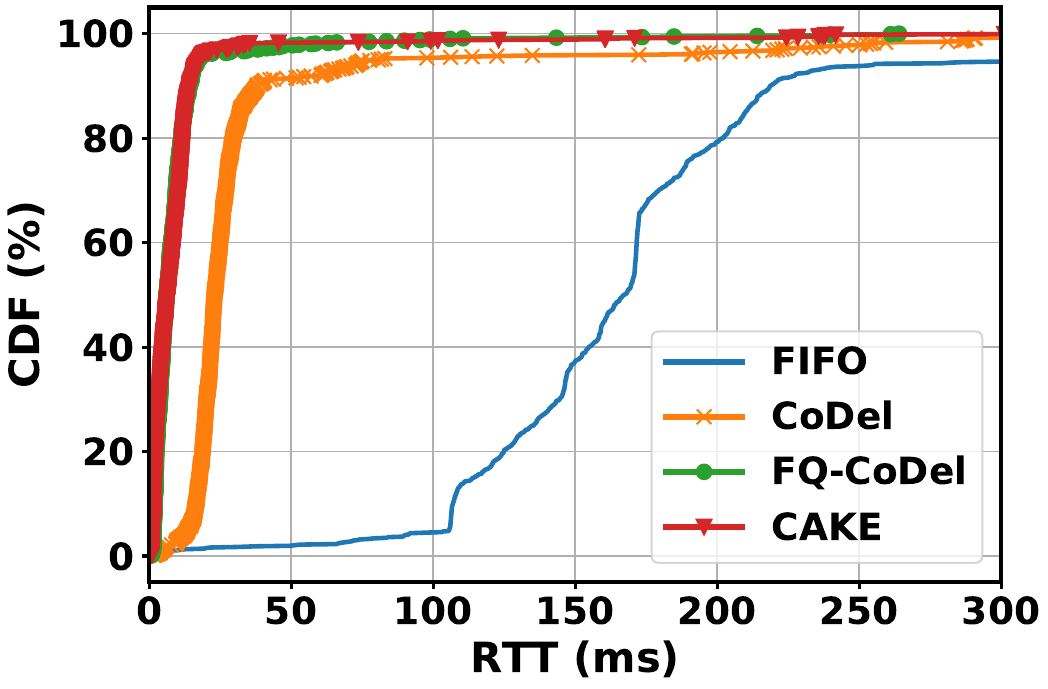}
    \caption{\Rmin{RTT (congested)}} \label{fig:CDFAIVrtt4}
\end{subfigure}\hfill%
\begin{subfigure}{.25\textwidth}
    \includegraphics[width=\linewidth]{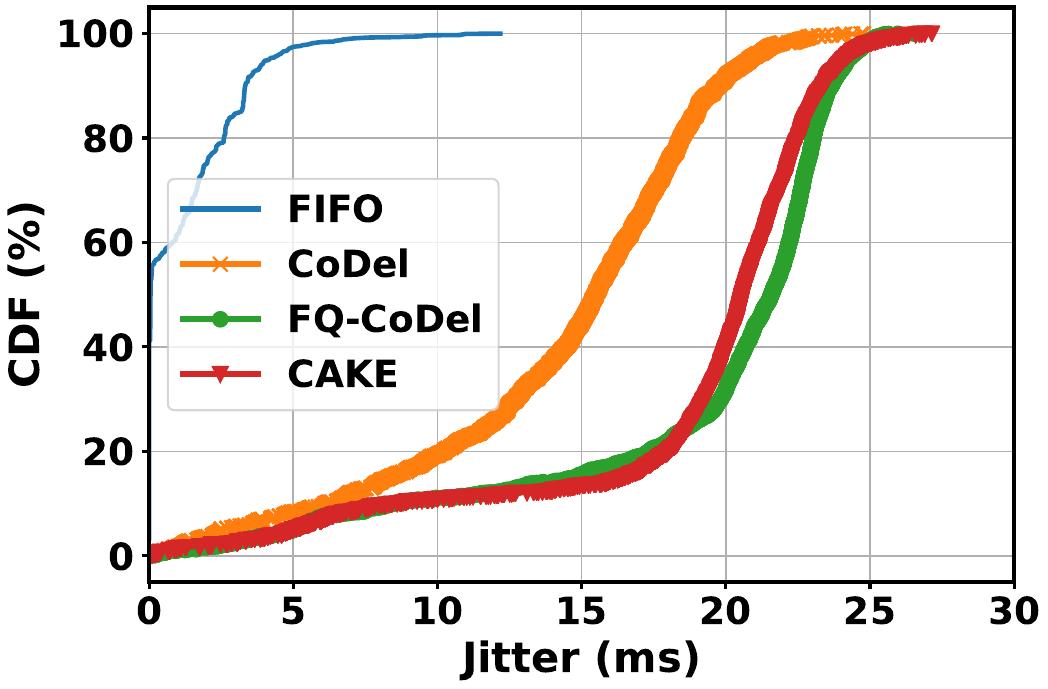}
    \caption{\Rmin{Jitter (congested)}} \label{fig:CDFAIVjitter4}
\end{subfigure}\hfill%
\begin{subfigure}{.25\textwidth}
    \includegraphics[width=\linewidth]{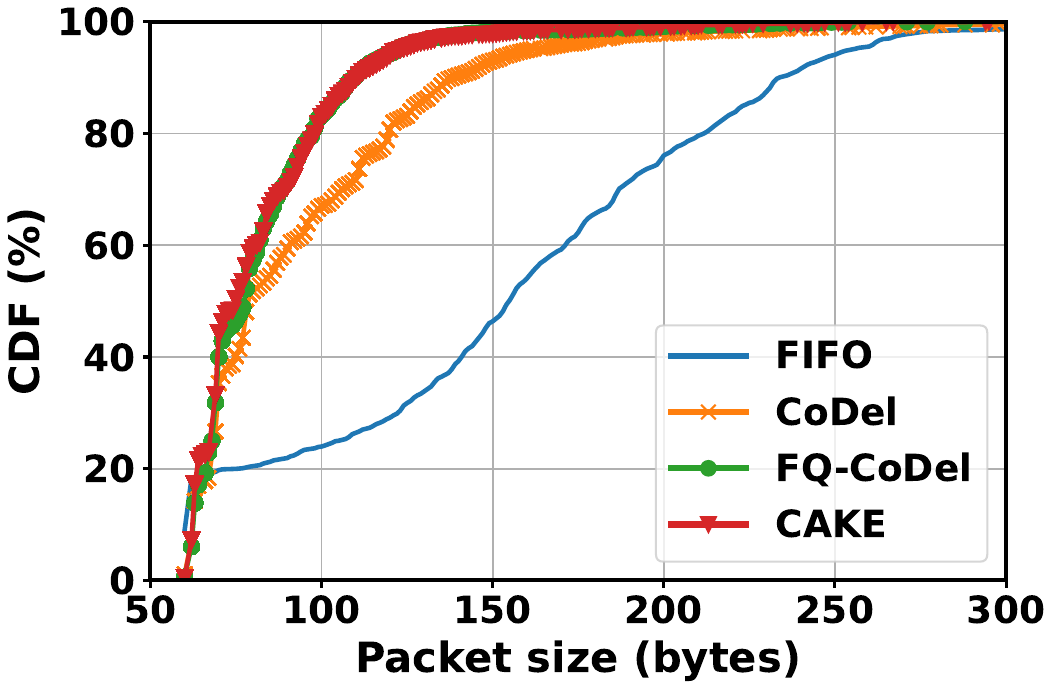}
    \caption{\Rmin{Packet size (congested)}} \label{fig:CDFAIVpacketdistr4}
\end{subfigure}\hfill%
\caption{\Rmin{CDF plots of throughput, RTT, jitter and packet size distribution of AIV1's traffic flow in uncongested (a)-(d) and congested (e)-(h) network conditions in Scenario II.}} \label{fig:CDF2}
\end{figure*}

\subsubsection{The impact on RTT and jitter} \label{subsubsection:jitter}
Fig.~\ref{fig:AIVtimeseriesRTT1} and~\ref{fig:AIVtimeseriesRTT2} show AIV1's induced RTT values across the experiment duration. In conjunction with the throughput performance, all queue disciplines demonstrate similar RTT values in the uncongested scenario, providing us with a baseline value of 10ms (on average). Deploying AQM in an uncongested (low bandwidth utilization) scenario does not provide any noticeable difference. However, when faced with the congestion (when bandwidth is highly utilised by other traffic), the benefit of AQM is evident. As demonstrated in Fig.~\ref{fig:AIVtimeseriesRTT2}, the RTT values in FIFO is inflated to 150-200ms, which is the typical bufferbloat phenomenon. When AQM is deployed, the RTT values are significantly lowered, with FQ-CoDel and CAKE providing the same RTT values (10ms) for the AIV (protecting AIV's flow) as the baseline. RTT values in CoDel is slightly higher (20ms) due to its single-queue management algorithm. Fig.~\ref{fig:AIVRTTBOX},~\ref{fig:CDFAIVrtt1} and~\ref{fig:CDFAIVrtt2} show the comparative boxplot and CDF plots for the induced RTT of AIV1's flow in both scenarios.


We note an important observation in the variation of packet jitter across all queue disciplines. \Rev{As shown in Fig.~\ref{fig:CDFAIVjitter1} and~\ref{fig:CDFAIVjitter2}, we observe that FQ-CoDel generally preserves the jitter characteristics most closely to the uncongested operational baseline across the evaluated scenarios. While the modified DRR scheduler (which is designed to serve a quantum of packets comprising 1500 bytes from each sub-queue in every iteration and prioritizes latency-sensitive flows, which is AIV1's flow in this case) likely contributes to this behavior by more evenly spacing packet transmissions across active flows, the observed jitter characteristics are also influenced by the interaction among per-flow queue isolation, queue management dynamics, traffic prioritization and congestion behaviors. We further note that CAKE also employs DRR-family scheduling and flow-isolation mechanisms, although differences in shaping behavior, queue management integration and traffic handling may contribute to the differences observed between CAKE and FQ-CoDel under congestion.}



\Rev{Although lower absolute jitter values are generally desirable from a networking perspective in the IT-domain, preserving timing characteristics close to the uncongested operational baseline is also important in the OT-domain. Lower absolute jitter values do not necessarily imply superior operational performance in industrial environments. For example, CAKE exhibits slightly lower absolute jitter distributions than FQ-CoDel in several congested scenarios. \Rmin{Nevertheless, preserving timing characteristics close to the normal operational baseline may be preferable for certain industrial applications.} Abrupt changes in packet timing behavior under congestion may affect synchronization stability, control responsiveness or fail-safe operation (even when the absolute jitter values are lower). The observed jitter behavior is influenced not only by DRR scheduling, but also by the interaction among queue isolation, shaping behavior, traffic prioritization and flow dynamics under congestion.}

\Rev{Therefore, jitter performance among FIFO, CoDel, FQ-CoDel and CAKE should be interpreted in the broader context of operational timing stability rather than solely on absolute jitter values. Due to this reason, our analysis considers not only absolute jitter values, but also the extent to which the AQM schemes preserve stable and predictable timing behavior relative to the uncongested baseline.}

\subsubsection{The impact on packet size distribution}

Fig.~\ref{fig:CDFAIVpacketdistr1} and~\ref{fig:CDFAIVpacketdistr2} investigate the packet size generated by AIV1 across all queue disciplines in both uncongested and congested scenarios. As the ground truth, all queue disciplines show similar packet size distribution (60 bytes, median), demonstrating the regular operation of AIV1 in the presence of video stream traffic. In the presence of competing elastic TCP flows, we observe the packet sizes generated in the FIFO scenario is unusually large (150 bytes, median). A closer packet inspection notes that the underlying transport protocol of AIV1 has clustered multiple payloads due to the extremely long delays incurred by bufferbloat in the bottleneck router (\textit{cf.} Fig.~\ref{fig:CDFAIVrtt2}). \Rmin{The internal mechanism of the transport protocol is designed conserve bandwidth utilization in low-bandwidth scenario, hence combining packet payloads while waiting for acknowledgment. However, this mechanism is detrimental to the performance of AIVs (and all IIoT flows, by extension) due to their requirement for timely data transmission for real-time operation. The adverse impacts are observed in our experiments where AIV1 enters the `fail-safe'/system malfunction mode.}

\subsection{Scenario II: Industrial automation network with real-time video streaming}

\Rmin{In Scenario II, we added an IP camera for real-time video streaming (used for surveillance and monitoring) to the network in the Cyber-Physical Factory cell. The IP camera is variable bitrate (VBR)-encoded, and generates a 7Mbps data rate on average~\cite{kua2017survey}.} \Rmin{We study the impact of such bandwidth-intensive application on industrial traffic flows with FIFO and AQM schemes, with/without the presence of elastic TCP flows.}

\subsubsection{The impact on throughput}
\Rmin{Fig.~\ref{fig:AIVtimeseriesTP3} and~\ref{fig:AIVtimeseriesTP4} demonstrate AIV1's throughput with the presence of IP camera traffic in both uncongested and congested scenarios. Since the bottleneck bandwidth is set to \Rmin{10 Mbps}, the VBR-encoded video stream generates \Rmin{7 Mbps} of throughput, and the IIoT flows are in the order of Kbps, all queue disciplines manage to handle the capacity sharing relatively well without the presence of elastic TCP flows.} Mission-critical flows are not severely impacted. However, the performance degradation with FIFO and the benefits of AQM schemes are clearly evident with the presence of elastic TCP flows. Fig.~\ref{fig:AIVtimeseriesTP4} shows that AIV1's throughput drastically drops to $<$ \Rmin{2 Kbps}, resulting in AIV1 stalling and going into `fail-safe' mode. Throughput with CoDel is much better at \Rmin{6 Kbps} (due to single-queue AQM), however it still falls short of the bitrate requirement for a functional AIV. FQ-CoDel and CAKE both provide the best throughput performance, preserving AIV1's throughput to the baseline in the presence of competing elastic flows. Fig.~\ref{fig:AIVflowtpBOX2},~\ref{fig:CDFAIVthroughput3} and~\ref{fig:CDFAIVthroughput4} present the comparison plots of AIV1's throughput in both scenarios, and clearly demonstrate the superiority of FQ-CoDel and CAKE.

\subsubsection{The impact on RTT and jitter}
Fig.~\ref{fig:AIVtimeseriesRTT3} and~\ref{fig:AIVtimeseriesRTT4} presents AIV1's induced RTT values with the presence of video streams, in uncongested and congested scenarios. As with the throughput performance, RTT values are similar across all queue disciplines in the uncongested scenario due to the sufficiently large bandwidth. However, the bufferbloat phenomenon is once again evident in the congested scenario, with RTT in FIFO spiking to 150ms on average (i.e., the RTT induced by full buffer capacity), rendering AIV1 unusable (\textit{cf.} Fig.~\ref{fig:AIVtimeseriesTP4}). CoDel, FQ-CoDel and CAKE provide significant latency improvements, with FQ-CoDel and CAKE performing the best (due to their set hashing-enabled capacity sharing abilities), hence allowing AIV1's induced RTT to remain at 10ms (as the baseline) and function as normal. Fig.~\ref{fig:AIVRTTBOX2},~\ref{fig:CDFAIVthroughput4} and~\ref{fig:CDFAIVrtt4} compare the induced RTT results and demonstrate the benefits of FQ-CoDel and CAKE. \Rev{Similar to the observations in Scenario I, we note that FQ-CoDel's per-flow queue isolation reduces interference between elastic and latency-sensitive traffic, hence preserving operational timing stability under congestion. It preserves packet jitter values (as demonstrated in Fig.~\ref{fig:CDFAIVjitter3} and~\ref{fig:CDFAIVjitter4}) due to its implementation of the modified DRR scheduler.} The modified DRR scheduler evenly spaces out packet transmission on the router's egress ports with an independently-managed sub-queue AQM algorithm. \Rev{In-depth analysis and insights into reasons (and broader context implications) behind the jitter variations in FQ-CoDel and CAKE are presented in \ref{subsubsection:jitter}}.

\subsubsection{The impact on packet size distribution}
Fig.~\ref{fig:CDFAIVpacketdistr3} and~\ref{fig:CDFAIVpacketdistr4} provide insights into the AIV1's packet size distribution as impacted by the queue disciplines in both uncongested and congested scenarios. As the baseline, all queue disciplines result in nearly identical (median) value of $<$ 60 bytes, representing the periodic and timely transmission of packet payloads. Under congestion, FQ-CoDel and CAKE preserve the distribution most closely to the baseline value, whereas CoDel's values skews slightly to the right, and FIFO shows the worst performance in a heavily right-skewed CDF plot. Upon inspection, it is noted at the transport stack of AIV1 combines multiple payloads into a single data packet due to the extremely high delay in waiting for the packet acknowledgment (rationale as discussed in Scenario I's analysis).


\subsection{\R{Scenario III: A hybrid wired/wireless multi-cell industrial automation network with real-time smart inventory and smart energy monitoring systems}}

\JK{Scenario III represents a sophisticated smart manufacturing scenario which implements a separate \textit{Additive Manufacturing cell} (located in a different physical location on the factory floor), in addition to the \textit{Cyber-Physical Factory cell} represented in Scenarios I and II. This scenario integrates all end-nodes, middleware and industrial network elements depicted in Fig.~\ref{fig:FoFtestbedtopology_both}. As described in Section~\ref{label:experimental_testbed_setup}, the separate Additive Manufacturing cell provides real-time inventory of raw materials with Gantry/BaSyX and smart energy monitoring, which informs the operation and optimization of the manufacturing process in the Cyber-Physical Factory cell (with AIVs handling the parts during the manufacturing process autonomously) via the BaSyX system and MES.}

\R{The real-time smart inventory system monitors and manage the availability of raw materials for 3D printing. The live visualization of data storage capacity is used to record past designs and predict material usage. When aligned with MES, the inventory system is able to optimize production schedules, optimize material usage and minimize waste. The real-time energy monitoring system tracks the energy usage of 3D printers, which enables the system to identify processes that consume large amounts of energy and subsequently reduce operational costs. Both smart inventory and energy monitoring systems represents real-time systems that are crucial for modern smart manufacturing to self-regulate and cooperate with different factory cells for an optimized production process.}

\JK{Traffic flows from both smart inventory and smart meter systems are generated by the OPC UA protocol to enable system-wide subscription of the live data, and to enable the interoperability of data transmission across industrial devices manufactured by different vendors. Both flows have different traffic characteristics. The application bitrate of the smart energy meter flow is \Rmin{25 Kbps}, whereas the bitrate of Gantry's operation (generated from the OPC UA server enabled within the Omron PLC) is \Rmin{2 Kbps}. As illustrated in Fig.~\ref{fig:FoFtestbedtopology}, the router connecting the entire Additive Manufacturing cell to the bottleneck router is implemented across a wireless link. Although the bitrates of both flows are relatively low, they are classified as mission-critical traffic flows due to their stringent latency requirements for real-time information exchange to/from the BaSyX system (located at the Edge Cloud level) via a hybrid wired/wireless communication infrastructure.}

\Rmin{In summary, this scenario specifically (i) investigates the impact of FIFO and AQM schemes on OPC UA-based mission-critical flows (for smart inventory and real-time energy monitoring); (ii) present the benefits of AQM in a hybrid wired/wireless multi-cell industrial automation network; and (iii) demonstrate the scalability of deploying AQM schemes across multi-cell factory floors, while accommodating the heterogeneity of low-rate mission-critical IIoT traffic flows.}

\subsubsection{The impact on throughput}
Fig.~\ref{fig:CDFgantrythroughput1},~\ref{fig:CDFgantrythroughput2},~\ref{fig:CDFenergythroughput1} and~\ref{fig:CDFenergythroughput2} present the throughput CDF distribution of Gantry and smart meter traffic flows in uncongested and congested environments, respectively. It is noted that the throughput performance of the traffic flows within the Cyber-Physical Factory cell are consistent with experiments conducted in Scenarios I and II. Here we analyze the throughput of the OPC UA flows within the Additive Manufacturing cell. \Rmin{Fig.~\ref{fig:scenario3boxplot} shows the comparison throughput and RTT boxplots of all three mission-critical traffic flows -- AIV, Gantry and smart meter traffic flows. The throughput boxplot of Gantry's flows are omitted due to their low values, but presented as a CDF plot in Fig.~\ref{fig:CDFgantrythroughput1} and~\ref{fig:CDFgantrythroughput2}.} Both Gantry and smart meter flows perform consistently well without congestion across FIFO, CoDel, FQ-CoDel and CAKE AQM. Gantry's throughput are not severely impacted under congestion due to its sporadic low-bitrate traffic of \Rmin{2 Kbps}.

However, smart meter flow across FIFO suffered significantly under congestion, with a drop of 40\% in throughput. While CoDel performs better in maintaining the median throughput values, it has a large variation due to its single-queue AQM characteristics. FQ-CoDel and CAKE demonstrates the best performance in preserving the throughput of the smart meter flow under congestion due to their flow-isolation and bandwidth sharing capabilities. The ability to preserve OPC UA flows' throughput is important in ensuring both Gantry and smart meter transmit intact information with the BaSyX system in real-time.

\subsubsection{\R{The impact on RTT and jitter}}

\begin{figure}
\centering
\begin{subfigure}{\linewidth}
    \centering
    \includegraphics[width=0.8\linewidth]{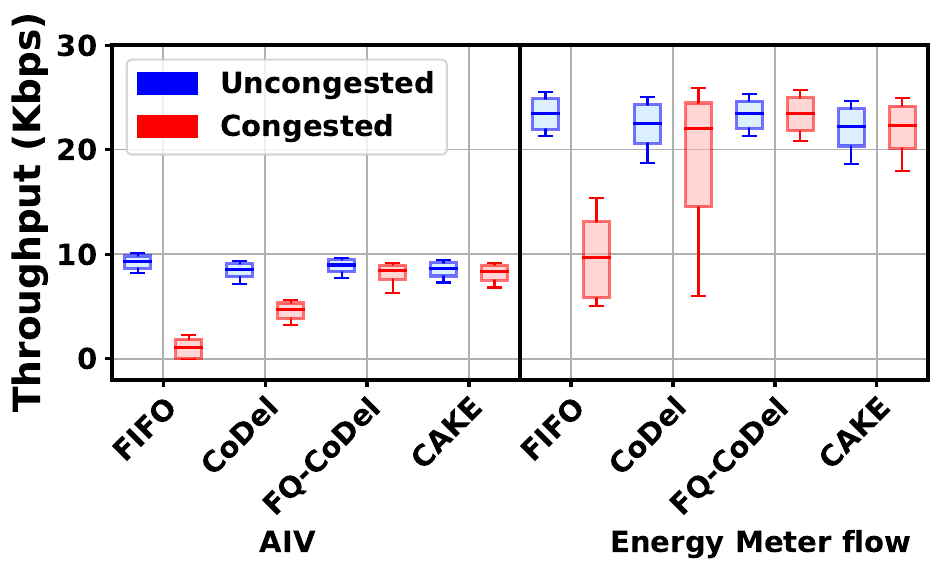}
    \caption{\Rmin{Throughput distribution}} \label{fig:missioncriticalthroughput}
\end{subfigure}\hfill%

\begin{subfigure}{\linewidth}
    \centering
    \includegraphics[width=0.8\linewidth]{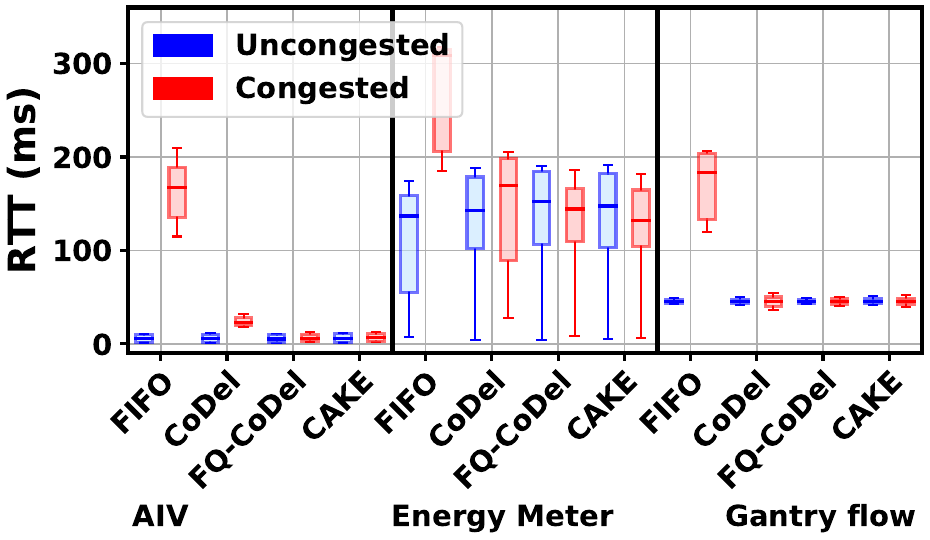}
    \caption{\Rmin{RTT distribution}}
    \label{fig:missioncriticalrtt}
\end{subfigure}\hfill%

\caption{\Rmin{Boxplots of throughput and RTT performance of AIV, smart energy meter and Gantry's traffic flows in uncongested and congested network conditions in Scenario III.}}
\label{fig:scenario3boxplot}
\end{figure}

\begin{figure*}[t]%
\centering
\begin{subfigure}{.25\textwidth}
    \includegraphics[width=\linewidth]{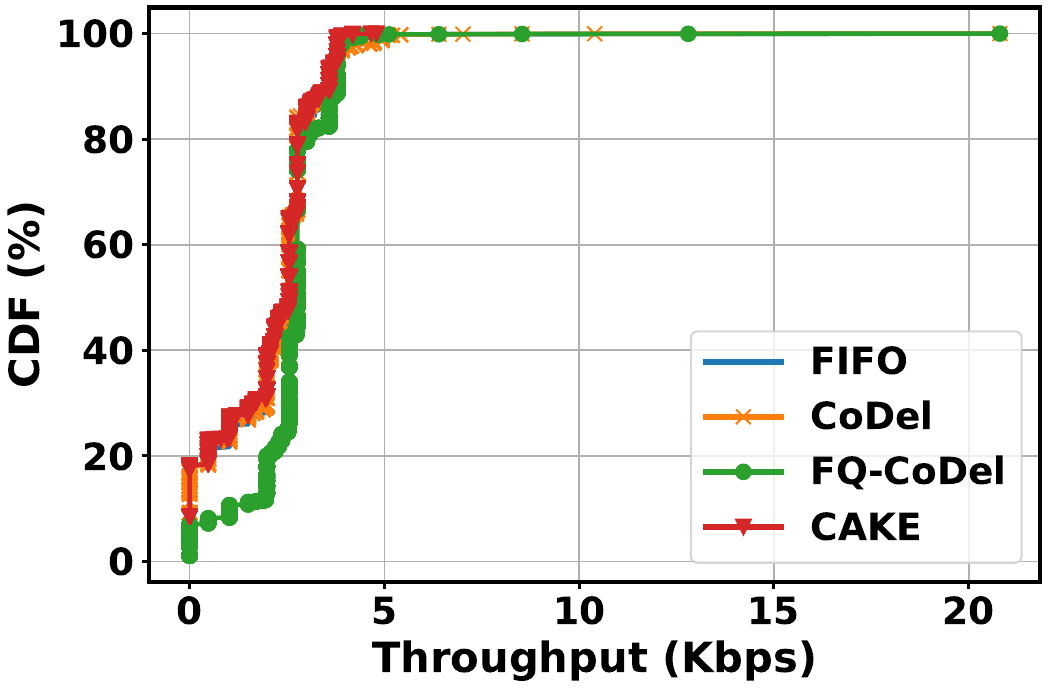}
    \caption{\Rmin{Throughput (uncongested)}} \label{fig:CDFgantrythroughput1}
\end{subfigure}\hfill%
\begin{subfigure}{.25\textwidth}
    \includegraphics[width=\linewidth]{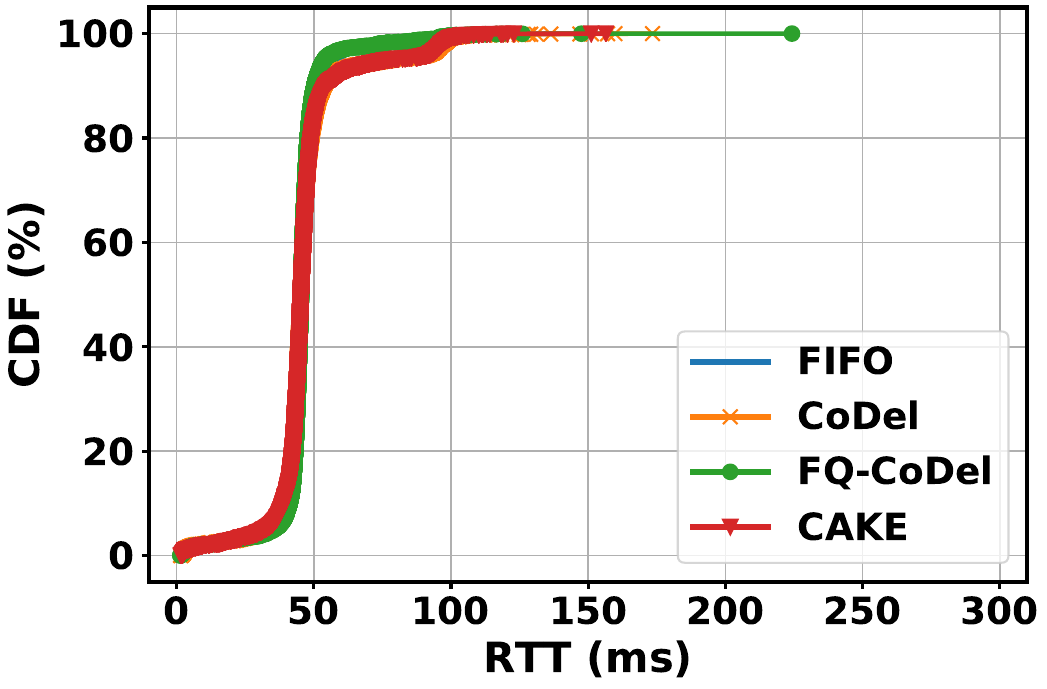}
    \caption{\Rmin{RTT (uncongested)}} \label{fig:CDFgantryrtt1}
\end{subfigure}\hfill%
\begin{subfigure}{.25\textwidth}
    \includegraphics[width=\linewidth]{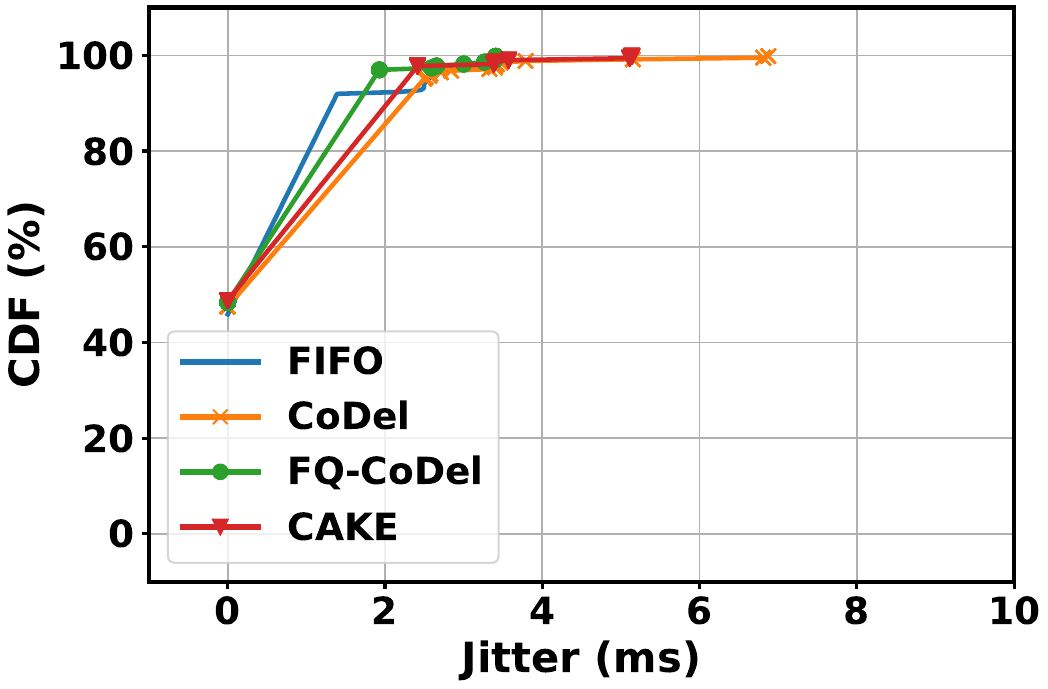}
    \caption{\Rmin{Jitter (uncongested)}} \label{fig:CDFgantryjitter1}
\end{subfigure}\hfill%
\begin{subfigure}{.25\textwidth}
    \includegraphics[width=\linewidth]{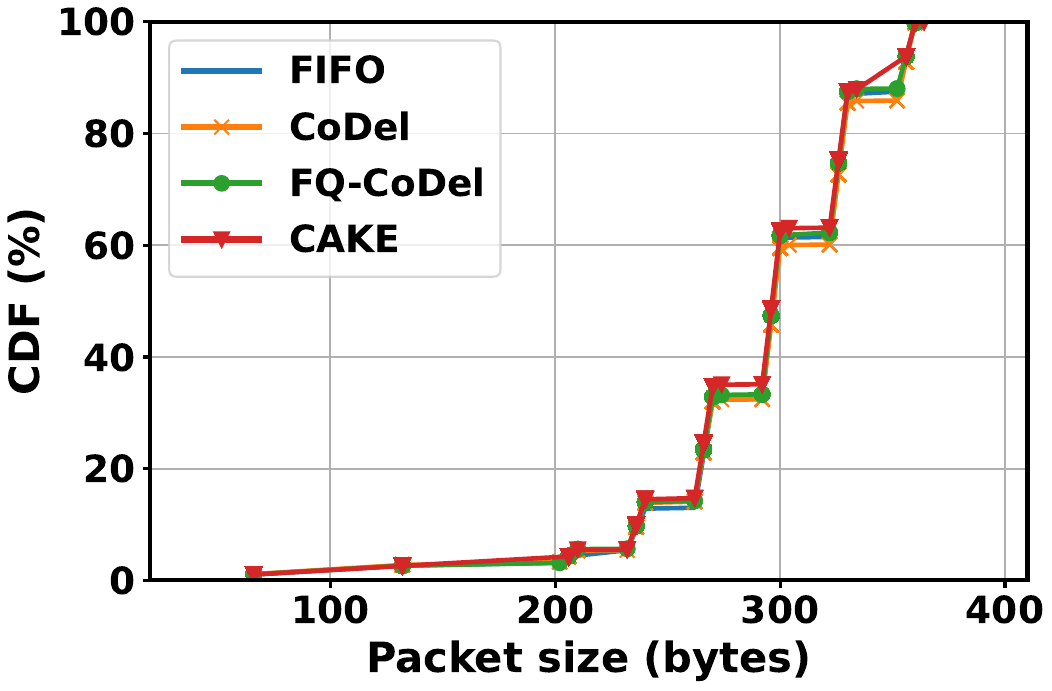}
    \caption{\Rmin{Packet size (uncongested)}} \label{fig:CDFgantrypacketdistr1}
\end{subfigure}\hfill%
\begin{subfigure}{.25\textwidth}
    \includegraphics[width=\linewidth]{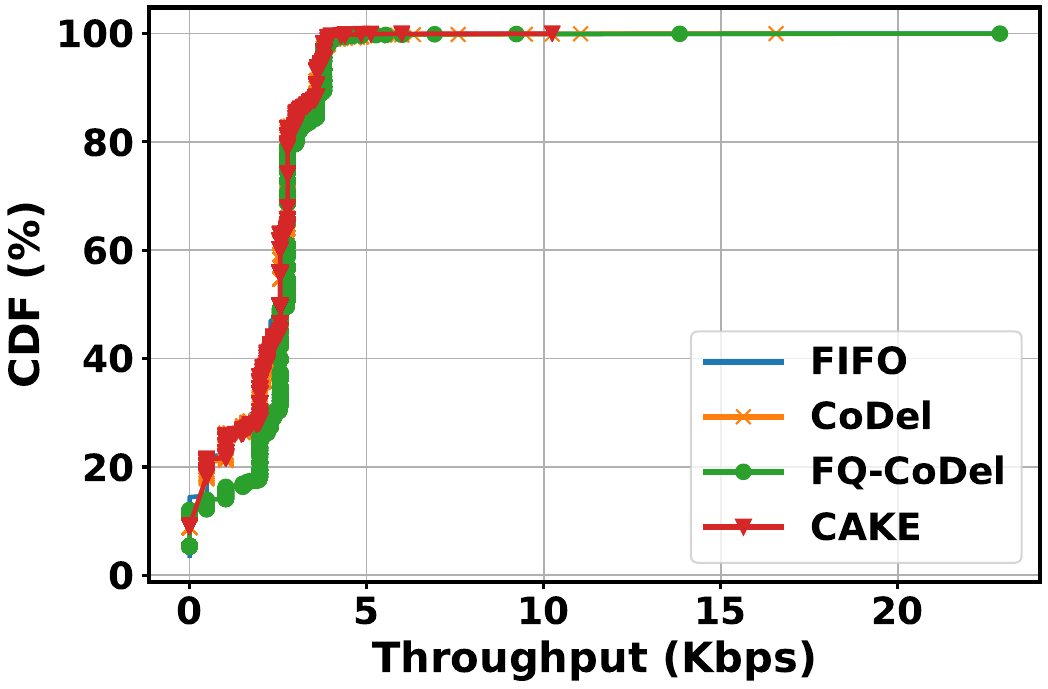}
    \caption{\Rmin{Throughput (congested)}} \label{fig:CDFgantrythroughput2}
\end{subfigure}\hfill%
\begin{subfigure}{.25\textwidth}
    \includegraphics[width=\linewidth]{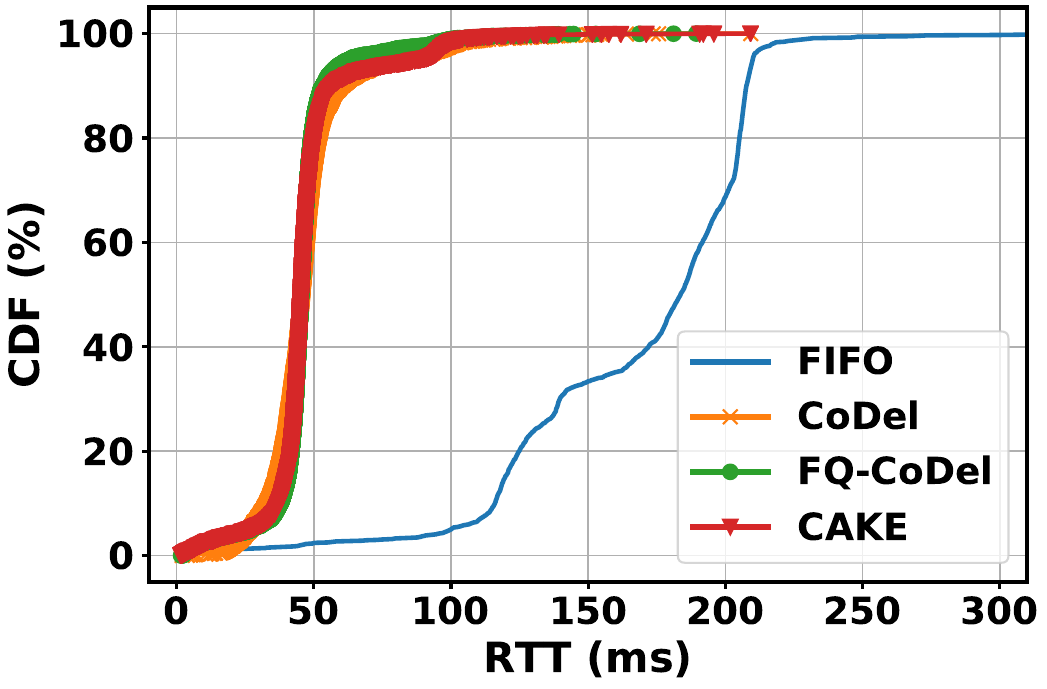}
    \caption{\Rmin{RTT (congested)}} \label{fig:CDFgantryrtt2}
\end{subfigure}\hfill%
\begin{subfigure}{.25\textwidth}
    \includegraphics[width=\linewidth]{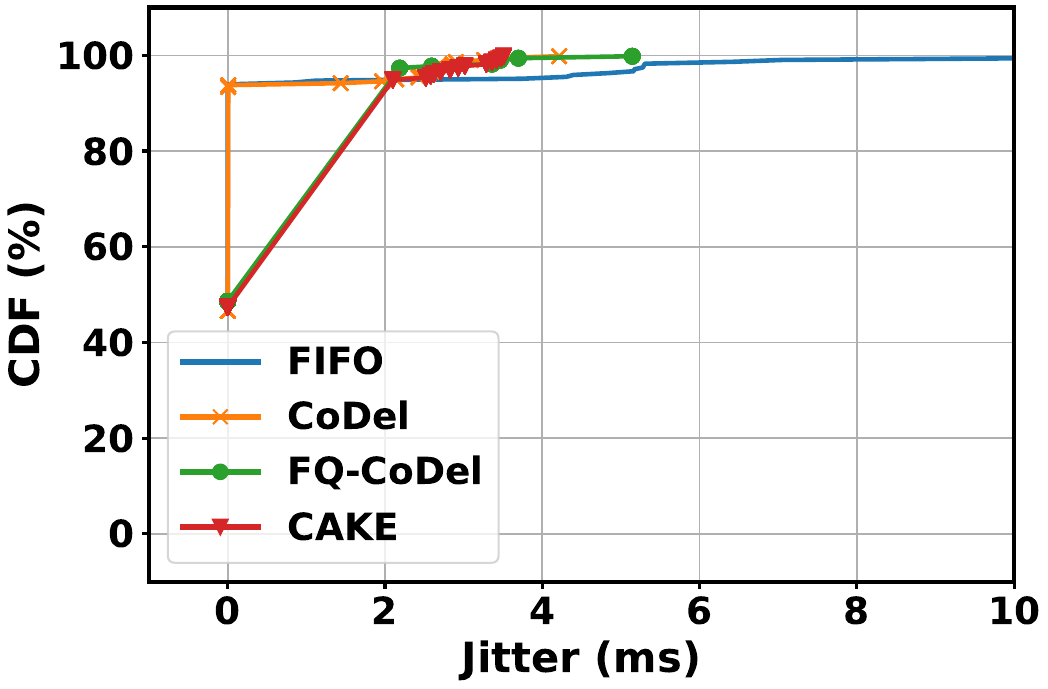}
    \caption{\Rmin{Jitter (congested)}} \label{fig:CDFgantryjitter2}
\end{subfigure}\hfill%
\begin{subfigure}{.25\textwidth}
    \includegraphics[width=\linewidth]{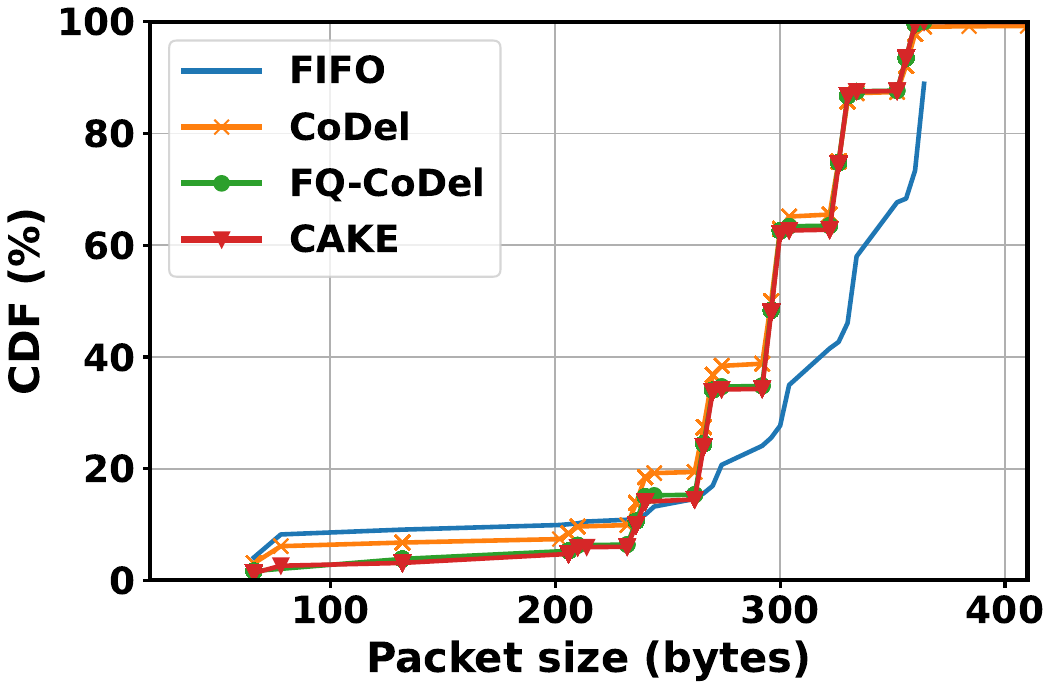}
    \caption{\Rmin{Packet size (congested)}} \label{fig:CDFgantrypacketdistr2}
\end{subfigure}\hfill%
\caption{\Rmin{CDF plots of throughput, RTT, jitter and packet size distribution of Gantry's OPC UA traffic flow in uncongested (a)-(d) and congested (e)-(h) experimental trials in Scenario III.}} \label{fig:gantryCDF2}
\end{figure*}

\begin{figure*}[t]%
\centering
\begin{subfigure}{.25\textwidth}
    \includegraphics[width=\linewidth]{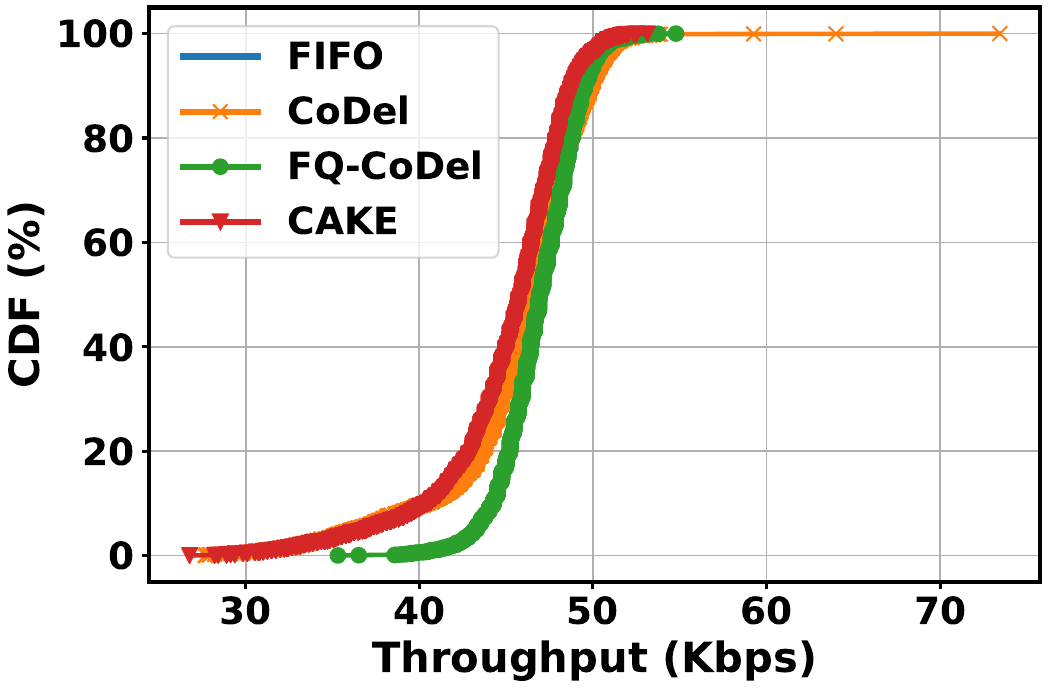}
    \caption{\Rmin{Throughput (uncongested)}} \label{fig:CDFenergythroughput1}
\end{subfigure}\hfill%
\begin{subfigure}{.25\textwidth}
    \includegraphics[width=\linewidth]{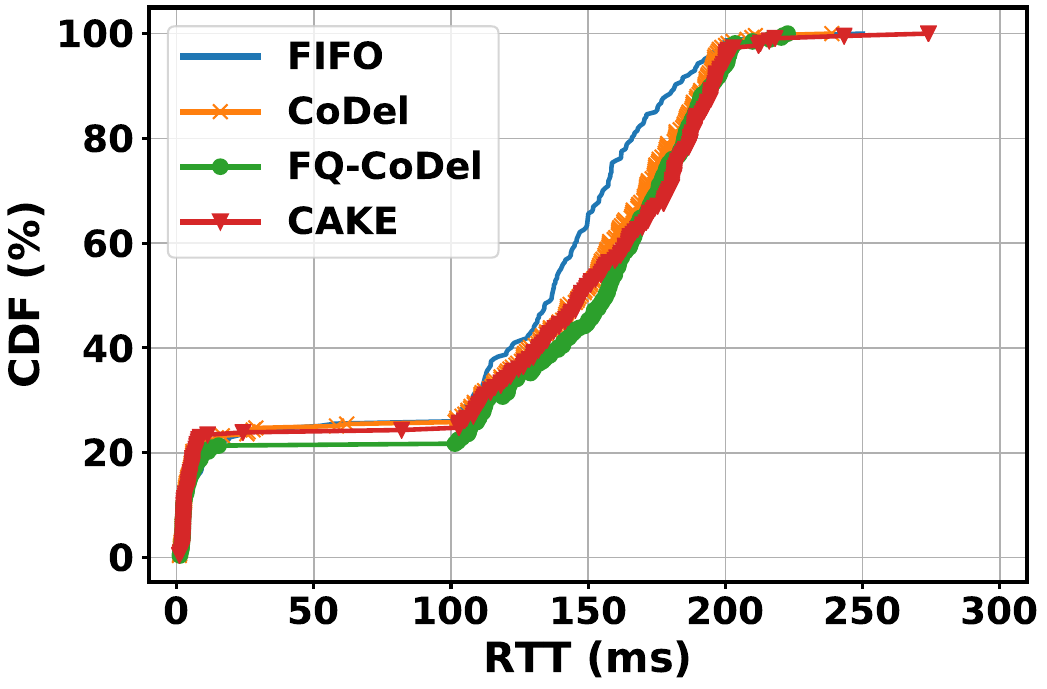}
    \caption{\Rmin{RTT (uncongested)}} \label{fig:CDFenergyrtt1}
\end{subfigure}\hfill%
\begin{subfigure}{.25\textwidth}
    \includegraphics[width=\linewidth]{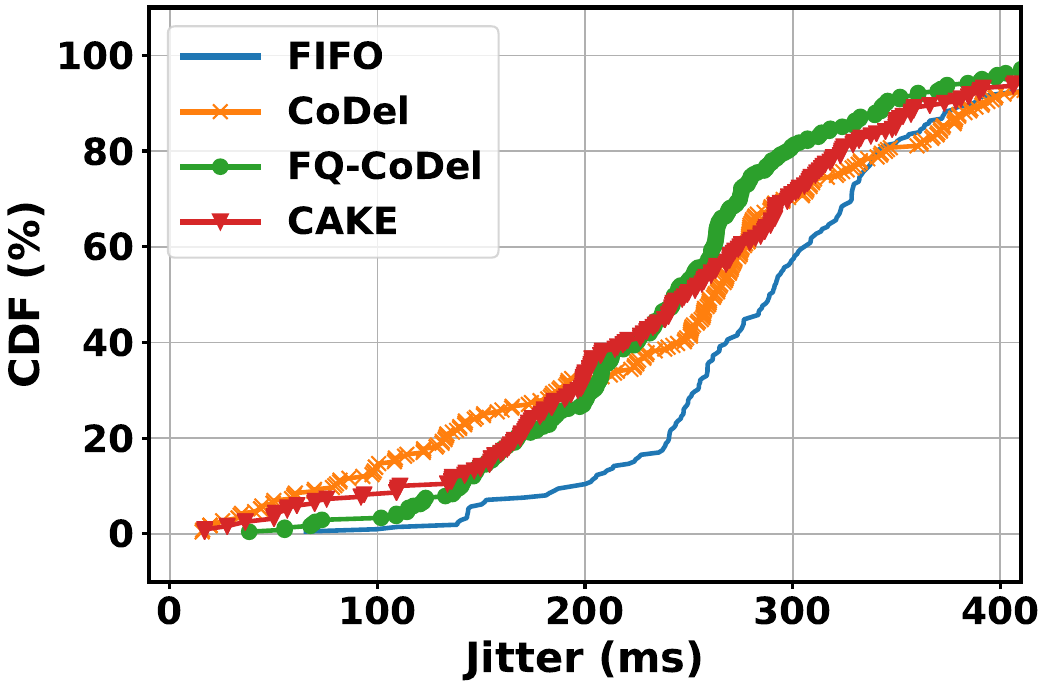}
    \caption{\Rmin{Jitter (uncongested)}} \label{fig:CDFenergyjitter1}
\end{subfigure}\hfill%
\begin{subfigure}{.25\textwidth}
    \includegraphics[width=\linewidth]{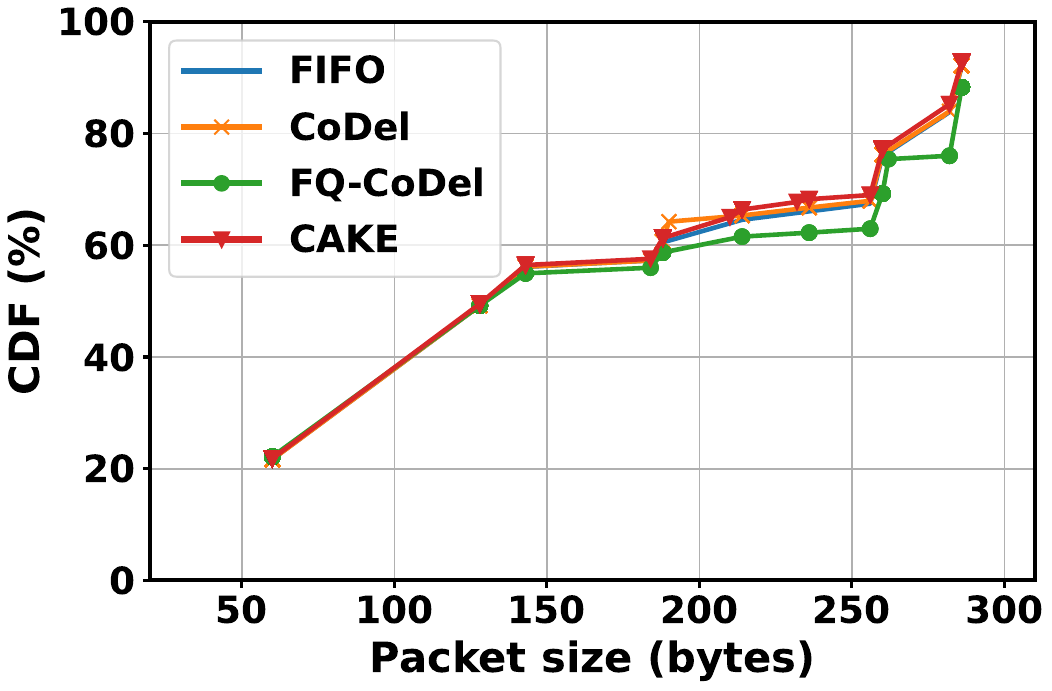}
    \caption{\Rmin{Packet size (uncongested)}} \label{fig:CDFenergypacketdistr1}
\end{subfigure}\hfill%
\begin{subfigure}{.25\textwidth}
    \includegraphics[width=\linewidth]{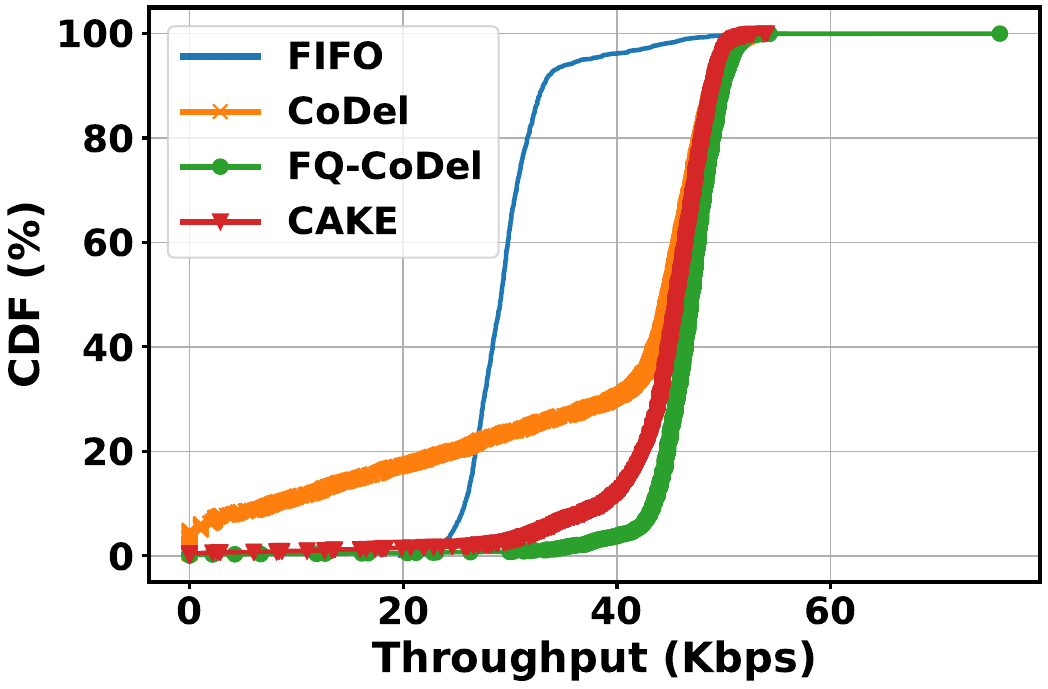}
    \caption{\Rmin{Throughput (congested)}} \label{fig:CDFenergythroughput2}
\end{subfigure}\hfill%
\begin{subfigure}{.25\textwidth}
    \includegraphics[width=\linewidth]{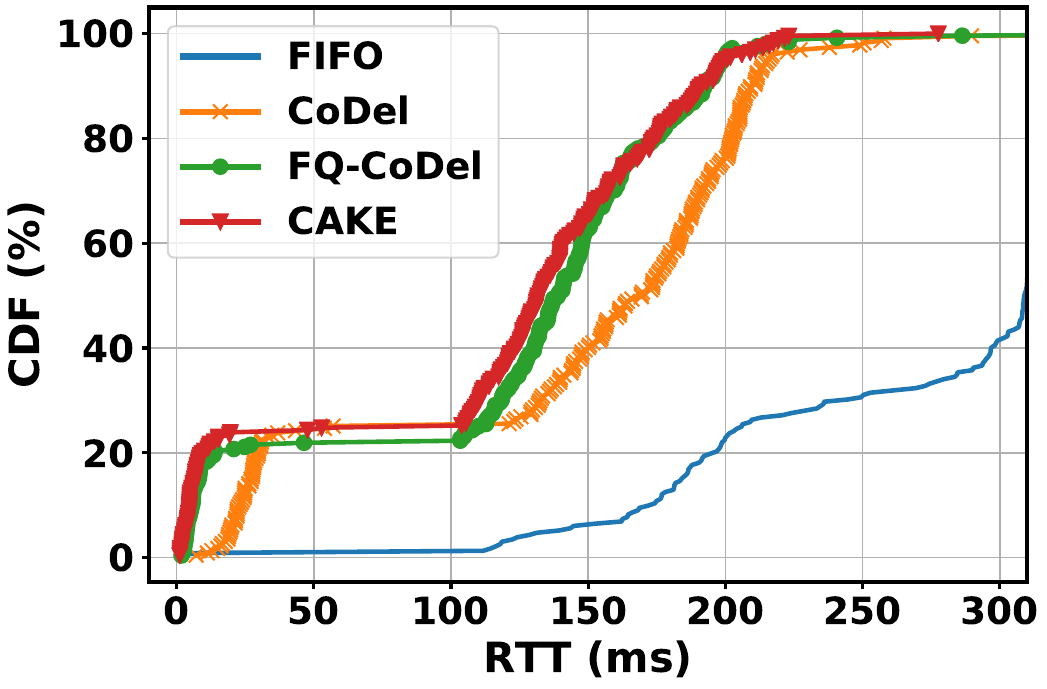}
    \caption{\Rmin{RTT (congested)}} \label{fig:CDFenergyrtt2}
\end{subfigure}\hfill%
\begin{subfigure}{.25\textwidth}
    \includegraphics[width=\linewidth]{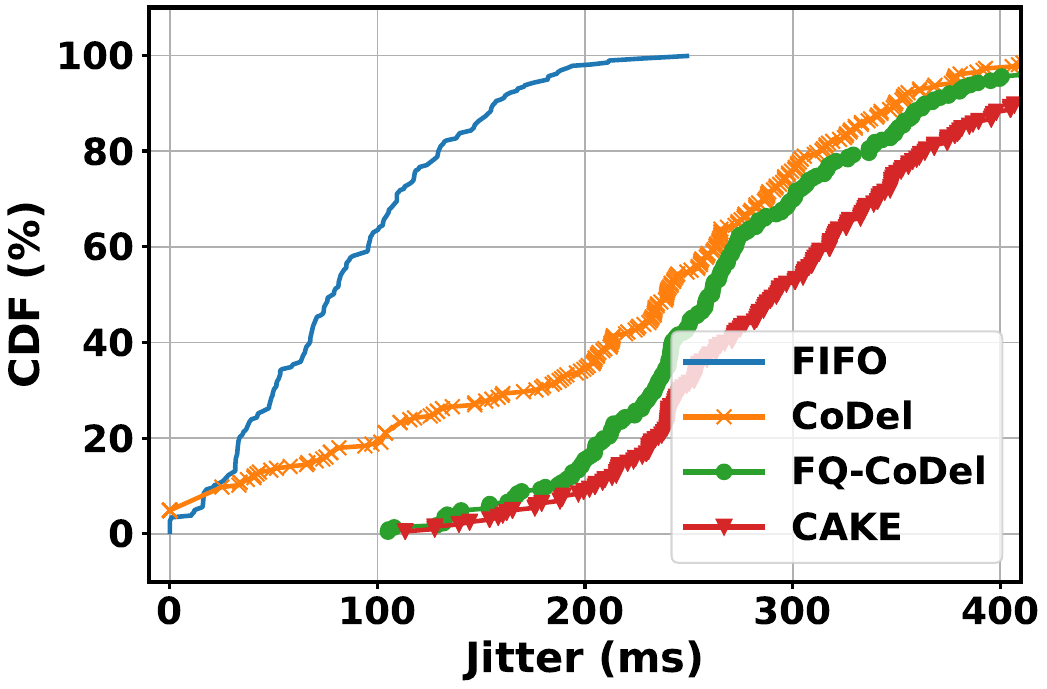}
    \caption{\Rmin{Jitter (congested)}} \label{fig:CDFenergyjitter2}
\end{subfigure}\hfill%
\begin{subfigure}{.25\textwidth}
    \includegraphics[width=\linewidth]{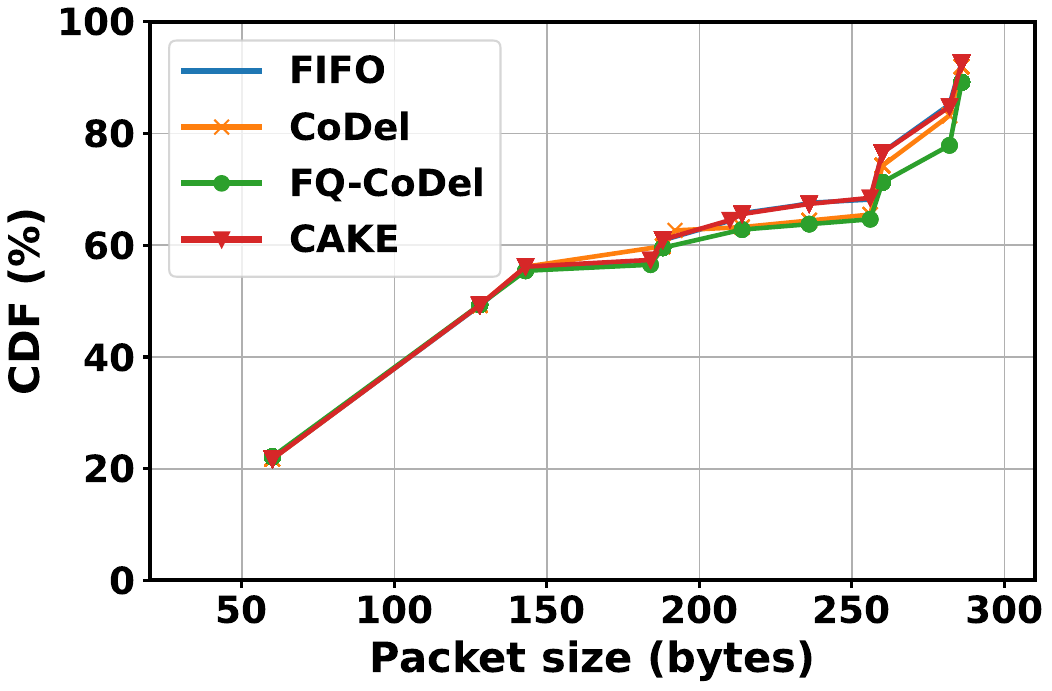}
    \caption{\Rmin{Packet size (congested)}} \label{fig:CDFenergypacketdistr2}
\end{subfigure}\hfill%
\caption{\Rmin{CDF plots of throughput, RTT, jitter and packet size distribution of Siemens smart energy meter OPC UA traffic flow in uncongested (a)-(d) and congested (e)-(h) experimental trials in Scenario III.}} \label{fig:energymeterCDF2}
\end{figure*}

Fig.~\ref{fig:CDFgantryrtt1},~\ref{fig:CDFgantryrtt2},~\ref{fig:CDFenergyrtt1} and~\ref{fig:CDFenergyrtt2} present the RTT distribution of Gantry and smart meter flows in uncongested and congested networks respectively, while Fig.~\ref{fig:missioncriticalrtt} presents the comparison boxplots of RTT measurements for AIV, Gantry and smart meter flows. Fig.~\ref{fig:CDFgantryjitter1},~\ref{fig:CDFgantryjitter2},~\ref{fig:CDFenergyjitter1} and~\ref{fig:CDFenergyjitter2} demonstrate the corresponding jitter variations. It is clearly evident that both FQ-CoDel and CAKE provides the best latency-reducing performance for all mission-critical flows with their multi-queue AQM mechanisms, including the two low-rate OPC UA-based traffic flows. CoDel demonstrates RTT values lower than FIFO but still exhibits large variation due to its single-queue characteristics. \Rev{The jitter distributions in Fig.~\ref{fig:CDFgantryjitter2} and \ref{fig:CDFenergyjitter2} exhibit overlapping trends across the evaluated AQM schemes. While FQ-CoDel and CAKE generally demonstrate more stable timing behavior under congestion, the relative ordering of the jitter distributions varies across different portions of the CDFs. This behavior reflects the complex interaction among queue buildup dynamics, flow isolation, scheduling behavior, shaping mechanisms and traffic characteristics under congested conditions. Hence, the results should not be interpreted independently (as discussed in Section~\ref{subsubsection:jitter}, but rather as demonstrating the difference in operational timing characteristics among the AQM schemes.}


\begin{figure}
\centering

\includegraphics[width=0.8\linewidth]{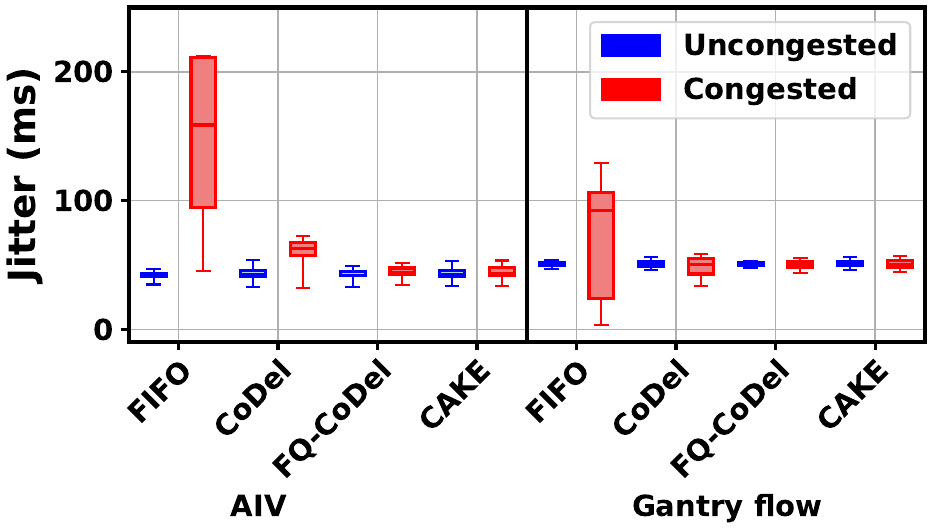}

\caption{\Rmin{Boxplot of jitter performance of AIV and Gantry's traffic flows in uncongested and congested network conditions in Scenario III.}}
\label{fig:scenario3_jitter_boxplot}
\end{figure}

\Rev{Fig.~\ref{fig:scenario3_jitter_boxplot} presents the boxplot of the jitter performance of AIV and Gantry's traffic flows. This complements the RTT analysis by showing the timing stability of the AIV and Gantry flows across different queue disciplines. In the uncongested case, jitter remains relatively stable for all schemes. \Rmin{Under congestion, FIFO exhibits the widest distribution and the largest timing variation, while CoDel provides only partial improvement. FQ-CoDel and CAKE consistently achieve the most compact jitter ranges, which indicates that their multi-queue flow-isolation mechanisms are more effective in protecting IIoT traffic from queue buildup and contention.} Although a left-shifted CDF plot can suggest lower absolute jitter, the boxplot shows if it comes with wider spread, longer whiskers or more variation. This observation supports the argument that stability and predictability are more important in IIoT traffic than lowering jitter to the smallest absolute value.
}

\Rev{Another observation is that although Gantry's throughput across FIFO remains at \Rmin{2 Kbps}, its RTT values are inflated to around \Rmin{200 ms} (300\% increase) which renders the flows unusable; whereas FQ-CoDel and CAKE preserve the RTT values of both Gantry and smart meter flows at \Rmin{50 ms and 150 ms} respectively (similar to when the network is uncongested), which satisfy their benchmarking requirements and translate into optimal performance in real-time data transmission to the BaSyX system. Therefore, the industrial significance of the observed jitter values should be interpreted in the context of timing stability and operational predictability rather than solely on absolute jitter values. Mission-critical industrial applications may tolerate small variations in instantaneous jitter, provided that the overall data communication behavior remains stable and predictable relative to the expected operational timing characteristics.}


\R{\subsubsection{The impact on packet size distribution}
Fig.~\ref{fig:CDFgantrypacketdistr1},~\ref{fig:CDFgantrypacketdistr2},~\ref{fig:CDFenergypacketdistr1} and~\ref{fig:CDFenergypacketdistr2} show the packet distribution of Gantry and smart meter flows in uncongested and congested environments, respectively. We observe a behavior in packet sizes in this scenario, which differs to Scenarios I and II. As demonstrated in the figures, the packet size distribution of Gantry and smart meter traffic flows are neither significantly impacted by network congestion nor the queue disciplines. Upon in-depth inspection of network and packet traces, we observed that the OPC server configuration within both PLCs has dedicated functional safety designs that ensure data packets of a specified length is consistently generated without relying on external acknowledgements, as opposed to TCP's behaviors (as observed and discussed in Scenarios I and II's analysis).}

\Rev{Our results consistently demonstrate the operational distinction between single-queue and multi-queue queue management approaches in IIoT. While CoDel effectively reduces excessive queue buildup compared to FIFO, the absence of explicit flow isolation limits its ability to protect latency-sensitive industrial traffic under sustained network congestion. In contrast, FQ-CoDel and CAKE are more effective in preserving timing stability and operational responsiveness by isolating competing traffic flows and reducing cross-traffic interference within the shared bottleneck/network infrastructure.}

\subsection{\Rev{Limitations}} \label{subsection:limitations}

\Rev{While our work provide useful operational insights into the impact of AQM schemes in the OT-domain/IIoT environments, we acknowledge several limitations and scope boundaries below:}

\begin{enumerate}[(i)]
    \item \Rev{The evaluated scenarios reflect representative Cloud-Fog Automation communication conditions using commodity networking infrastructure and heterogeneous industrial traffic flows. However, they do not exhaustively capture all industrial deployment scales, topologies or traffic characteristics. Larger deployments may exhibit additional communication heterogeneity, synchronization requirements, wireless interference dynamics and distributed edge-cloud coordination complexity.}
    \item \Rev{The generated traffic patterns emulate representative mission-critical industrial communication behavior, including OPC UA telemetry, robotic coordination traffic, supervisory communication and competing elastic traffic flows. Nevertheless, real industrial environments may involve additional proprietary protocols, deterministic communication schedules and application-specific traffic dynamics that can influence the performance of AQM schemes.}
    \item \Rev{Our study focuses on AQM behavior within conventional IP-based networking infrastructure and does not evaluate deterministic networking technologies such as TSN or DetNet. Hence, the results should not be interpreted as a replacement for deterministic industrial networking architectures, but rather as an investigation of lightweight and backward-compatible AQM approaches suitable for exisiting IIoT systems.}
    \item \Rev{Although our experiments demonstrate consistent operational trends across the evaluated scenarios, broader evaluation across larger industrial deployments, new wireless connectivity conditions and more diverse operational workloads remains important future work.}
\end{enumerate}

\Rev{Despite these limitations, the work presented in this paper demonstrate consistent operational trends across heterogeneous industrial communication scenarios and provide practical insights into the deployment implications of modern AQM schemes in for current/legacy IIoT systems transitioning towards a new Cloud-Fog Automation-based paradigm.}

\subsection{\JK{Practical deployment considerations and recommendations}}

\Rev{In this section, we provide several practical deployment considerations and recommendations for industry practitioners and researchers in the field.}

\subsubsection{\Rev{Hardware and infrastructure}}
\Rev{A practical advantage of deploying modern AQM mechanisms in industrial OT/IIoT environments is their compatibility with commodity IP-based networking infrastructure. Unlike deterministic networking approaches that may require specialized switching hardware and tightly coordinated scheduling mechanisms, AQM schemes such as CoDel, FQ-CoDel, and CAKE are widely supported in modern Linux-based networking stacks and commodity routing platforms. This enables incremental deployment within existing industrial communication infrastructure with relatively low hardware replacement cost.
}

\subsubsection{\Rev{Interoperability}}
\Rev{Interoperability remains an important consideration in industrial networking environments due to the coexistence of heterogeneous protocols, legacy automation systems, vendor-specific equipment, and mixed communication technologies. In this context, the evaluated AQM mechanisms operate transparently at the network and queue-management level without requiring modification to application-layer industrial protocols such as OPC UA or changes to the underlying industrial devices and PLCs. This backward-compatible behavior is particularly important in transitional Cloud-Fog Automation environments where complete infrastructure replacement may not be operationally or economically feasible.}

\subsubsection{\Rev{Deployment recommendations and caution}}
\Rev{Despite the observed benefits, operational deployment of AQM in industrial environments should be approached carefully. Industrial systems often contain tightly coupled control processes and timing-sensitive applications where their behavior may depend on stable communication characteristics. AQM configurations should be validated under representative operational traffic conditions prior to deployment, and conservative parameter selection should be preferred unless extensive scenario-specific evaluation has been conducted. In summary, we provide the following deployment considerations/recommendations based on our experimental insights:}
\begin{enumerate}[(i)]
    \item \JK{Modern IIoT facilities (e.g., smart factories) use IT-based routers as gateways/proxies for OT environments. Although AQM schemes such as PIE, CoDel, FQ-CoDel, CAKE are provided, FIFO remains the default queue discipline in most industrial routers. The trepidation to enable AQM in OT environment is the concern that AQM can cause unpredictable packet losses to OT flows. However, we have experimentally proven in this work that AQM does not adversely impact the performance of IIoT traffic. Rather, AQM schemes significantly improve throughput and RTT performance (especially mission-critical flows with stringent real-time requirements) in low-bandwidth and congested network environments.}
    \item \JK{Network performance is sensitive to the internal AQM parameters, as they dictate how the bottleneck queue signals and interacts with the end-to-end transport protocols for regulating data transmission. Global AQM parameters (which are extensively tested and intricately tuned) should be left as default and ECN should be disabled, unless extensive experiments have been conducted on scenario-specific network configurations.}
    \item \JK{Practitioners should be guided by RFC7567~\cite{rfc7567} and RFC7928~\cite{rfc7928} when deploying AQM schemes in industrial settings. AQM behaviors should be first assessed in a controlled environment (with the recommended traffic mix), to ensure a safe operational deployment. Practitioners should consider the potential impact of the central aspects of an AQM algorithm, such as their burst absorption capacity, RTT fairness, resilience to fluctuating network conditions, on the intended use cases.}
\end{enumerate}

\Rev{Although our experimental scenarios do not represent exhaustive large-scale industrial deployments, the observed behavior suggests that multi-queue AQM mechanisms remain beneficial in heterogeneous IIoT environments where mission-critical and elastic traffic co-exist over shared communication infrastructure.}

\Rev{As industrial deployments scale in terms of traffic diversity, subsystem heterogeneity and edge-cloud coordination complexity, flow-isolation and queue management behavior are expected to become increasingly important for preserving operational timing stability under network congestion.}

\section{Conclusions and Future Work}
\label{Section:Conclusions}


In this work, we proposed and experimentally studied the use of AQM schemes to assist mission-critical traffic flows in a real-world IIoT environment with three realistic network scenarios. There is currently a building momentum towards a more network-centric and collaborative paradigm, known as Cloud-Fog Automation. To progressively realize this paradigm with state-of-the-art networking technologies, operational networks in IIoT first require a backward-compatible and non-intrusive technique to alleviate existing network pain points.

\Rev{Our work demonstrates that the commonly-observed bufferbloat phenomenon in the IT-domain is also prevalent in the OT-domain. We implemented emerging AQM schemes (CoDel, FQ-CoDel and CAKE) in our network, and showed that AQM schemes benefit all OT traffic flows, especially in assisting mission-critical traffic flows to be unperturbed (and protected) in heavily congested environments. Both FQ-CoDel and CAKE are superior due to their sophisticated flow isolation and capacity sharing abilities.} We observed that FQ-CoDel and CAKE enable mission-critical flows to achieve up to 80\% performance improvement when compared with FIFO -- allowing the production process to continue seamlessly in spite of heavy competing traffic. Furthermore, we have demonstrated that AQM schemes can assist OPC UA flows in a sophisticated hybrid wired/wireless environment, which encourages the wider deployment of AQM for supporting protocol interoperability in IIoT networks.

\Rmin{The core contribution of this work is the experimental characterization of the operational benefits and impact of modern AQM schemes on the mission-critical industrial traffic performance in the OT-domain/IIoT environments (and not in proposing a new IIoT-specific AQM scheme, which is the next step in our future research). It is our sincere hope that this foundational work provides the motivation, experimental characterization and validation for operators to deploy AQM in modern industrial networks.}

\Rev{Future work include investigating the limitations outlined in Section~\ref{subsection:limitations}, and studying AQM parameter-tuning for IIoT environments, implementing AI/ML-driven decision-making into IIoT-based AQM schemes, validating protocol correctness, integrating AQM with in-network computing technologies to achieve network-wide deterministic low-latency networking. A rethink and redesign of queue management strategies is required to cater for non-TCP/IP, ultra-Reliable Low-latency Communication (uRLLC) traffic flows in emerging 5G/6G networks~\cite{buadoi2026information,zeydan20246g}.}


\printcredits

\bibliographystyle{model1-num-names}


\bibliography{IEEEabrv,references_aqm_iiot}

\end{document}